\documentclass[a4paper,11pt]{article}
\pdfoutput=1
\usepackage{jheppub}
\usepackage{wasysym}
\usepackage{amssymb,amsmath}
\usepackage{mathrsfs}
\usepackage{graphicx}
\usepackage{epsfig}
\usepackage{epstopdf}
\usepackage{latexsym}
\usepackage{psfrag}
\usepackage{enumitem}
\usepackage{booktabs}
\usepackage{bbm}

\usepackage[toc]{appendix}

\usepackage{color}
\usepackage[dvipsnames]{xcolor}
\usepackage{hyperref}
\usepackage[T1]{fontenc} % needed for Icelandic letters
\hypersetup{colorlinks,
	linkcolor = blue,
	anchorcolor = red,
	citecolor = blue,
	filecolor = red,
	urlcolor = blue,
	linktocpage}

\numberwithin{equation}{section}  %this is the correct way to number equations within sections
\renewcommand{\theequation}{\arabic{section}.\arabic{equation}}

\usepackage{benmacros}

\title{\boldmath Holographic dual of hot Polchinski-Strassler \\ quark-gluon plasma}
\author[1]{Iosif~Bena,}
\author[2]{\'Oscar~J.~C.~Dias,}
\author[2]{Gavin~S.~Hartnett,}
\author[3]{Benjamin~E.~Niehoff,}
\author[4]{Jorge~E.~Santos}
\affiliation[1]{Institut de Physique Th\'{e}orique, Universit\'{e} Paris Saclay, CEA, CNRS,\\ 91191 Gif-sur-Yvette Cedex, France}
\affiliation[2]{STAG research centre and School of Mathematical Sciences, University of Southampton,\\ Highfield Campus, Southampton SO17 1BJ, UK}
\affiliation[3]{Institute for Theoretical Physics, KU Leuven Celestijnenlaan 200D, B-3001 Leuven, Belgium}
\affiliation[4]{Department of Applied Mathematics and Theoretical Physics,\\University of Cambridge, Wilberforce Road, Cambridge CB3 0WA, UK}
\emailAdd{iosif.bena@cea.fr}
\emailAdd{ojcd1r13@soton.ac.uk}
\emailAdd{gshartnett@gmail.com}
\emailAdd{ben.niehoff@kuleuven.be}
\emailAdd{jss55@cam.ac.uk}

\abstract{We construct the supergravity dual of the hot quark-gluon plasma in the mass-deformed ${\cal N}=4$ Super-Yang-Mills theory (also known as ${\cal N}=1^*$). The full ten-dimensional type IIB holographic dual is described by 20 functions of two variables, which we determine numerically, and it contains a black hole with $S^5$ horizon topology.  As we lower the temperature to around half of the mass of the chiral multiplets, we find evidence for (most likely a first-order) phase transition, which could lead either to one of the Polchinski-Strassler confining, screening, or oblique vacua with polarized branes, or to an intermediate phase corresponding to blackened polarized branes with an $S^2 \times S^3$ horizon topology.  This phase transition is a feature that could in principle be seen by putting the theory on the lattice, and thus our result for the ratio of the chiral multiplet mass to the phase transition temperature, $m_c/T = 2.15467491205(6)$, constitutes the first prediction of string theory and AdS/CFT that could be independently checked via four-dimensional super-QCD lattice computation.  We also construct the black-hole solution in certain five-dimensional gauged supergravity truncations and, without directly using uplift/reduction formulae, we find strong evidence that the five- and ten-dimensional solutions are the same. This indicates that  five-dimensional gauged supergravity is powerful enough to capture the physics of the high-temperature deconfined phase of the Polchinski-Strassler quark-gluon plasma.

\bigskip

\bigskip

\centerline{\it Dedicated to the memory of Joe Polchinski}
}

\begin{document}
\maketitle

%%%%%%%%%%%%%%%%%%%%%%%%%%%%%%%%%%%%%
%%%%% SECTION %%%%%%%%%%%%%%%%%%%%%%%
\section{Introduction}

One of the hardest problems in theoretical particle physics is understanding confinement. It is well-known that the Standard Model Lagrangian gives a good description of quarks and gluons at high energies, but at lower energies Quantum Chromodynamics (QCD) becomes strongly coupled and is unable to describe the physics of the confined phase. The only two theoretical pathways for describing this physics are Lattice QCD and the AdS/CFT correspondence.

Each of these methods has its advantages. Lattice methods have been used for more than forty years, and for example can obtain the mass of the proton to a reasonably good precision. However, they cannot describe real-time dynamics of the quark gluon plasma observed at RHIC and LHC. On the other hand, the holographic AdS/CFT correspondence can be used to determine this dynamics, as well as many other features that cannot be captured by lattice calculations. Yet this method, too, is of limited applicability. In particular, it cannot be used to describe exactly the same theory as QCD, but rather `cousin theories' that one hopes to be in the same universality class. The theory that is easiest to study using the AdS/CFT correspondence is the maximally-supersymmetric ($\mathcal{N} = 4$) Super Yang Mills (SYM) theory with a large-$N$ $SU(N)$ gauge group, which is a conformal theory with matter in the adjoint representation and which does not exhibit confinement. Although $\mathcal{N}=4$ SYM is very different from the theory of QCD, it is able to capture certain experimentally observed phenomena of QCD reasonably well, such as the low viscosity-to-entropy ratio or jet quenching \cite{Policastro:2001yc,Kovtun:2004de}. 

However, there is a lot more physics to QCD than these phenomena, and most of this physics does not exist in a maximally-supersymmetric conformal field theory. To describe these phenomena holographically, one has to deform the AdS/CFT correspondence by breaking supersymmetry and conformal invariance. These deformations are well understood \cite{Banks:1998dd,Balasubramanian:1998de}: adding relevant deformations to the $\mathcal{N} = 4$ Lagrangian corresponds to turning on non-normalizable modes in the dual $AdS_5 \times S^5$ geometry. In general these modes change the geometry drastically, and the new geometry can yield valuable information about the physics of the deformed theory.

Of course, in order to get closer to real-world QCD one is most interested in deformations that render the infrared theory confining. For $\mathcal{N} = 4$ SYM these deformations consist of giving masses to the fermions and the scalars. In general these mass deformations break all supersymmetry, but there exist certain combinations that preserve four supercharges. The resulting theory, known as $\mathcal{N} = 1^*$, has three massive chiral multiplets and a vector multiplet. This theory was studied in great detail both using supersymmetric methods \cite{Donagi:1995cf}, and using the AdS/CFT correspondence \cite{Polchinski:2000uf}, and was found to have a very rich structure of supersymmetric vacua. Some of these vacua are confining, some are screening, and some correspond to so-called oblique phases in which only certain combinations of quarks and monopoles are screened. In the holographic dual, Polchinski and Strassler have argued \cite{Polchinski:2000uf} that the confining vacua correspond to asymptotically-$AdS_5 \times S^5$ solutions that have D3 branes polarized into NS5 branes in the infrared, the screening vacua (also known as the Higgs vacua) have D3 branes polarized into D5 branes, while the oblique vacua have D3 branes polarized into $(p, q)$ five-branes.\footnote{The polarization of the D3 branes (also known as the Myers effect \cite{Myers:1999ps}) is caused precisely by the supergravity fields corresponding to the fermion and boson masses.} In addition to these supersymmetric zero-temperature phases, this theory has a high-temperature phase, which is expected to be dual to an asymptotically $AdS_5 \times S^5$ solution with a black hole in the infrared \cite{Freedman:2000xb}, and possibly other finite-temperature phases that may be dominant and sub-dominant at intermediate temperatures.

There are several features that make this $\mathcal{N} = 1^*$ theory by far the best candidate for extending the holographic correspondence towards theories that are closer to QCD. The first is its rich phase structure, which allows one to study the confinement/deconfinement phase transition, the tunneling between different vacua, the melting of mesons and baryons, etc. The second is its supersymmetry, which ensures the absence of a Weyl anomaly, and hence the absence of fields that depend logarithmically on the radius and that could spoil the convergence of the numerical analysis. The third, and perhaps most important, is the fact that the UV of the theory is described by a well-defined four-dimensional Lagrangian, and therefore this theory may be amenable to lattice simulations. None of the other string theory solutions that have been proposed to describe the physics of the confined phase share this feature: their UV regions either correspond to some complicated ten-dimensional string-theory configuration (such as the Sakai-Sugimoto system \cite{Sakai:2004cn}), or to compactified higher-dimensional theories \cite{Witten:1997ep,Giveon:1998sr}, or to four-dimensional theories on a compact manifold \cite{Witten:1998zw}, or to theories whose gauge group rank keeps on running forever, and which correspond in the bulk to solutions with an infinite charge.\footnote{The Klebanov-Strassler solution \cite{Klebanov:2000hb} is the best known example in this class.}

Thus, the $\mathcal{N} = 1^*$ theory is the only known four-dimensional confining theory that has any chance to be studied simultaneously using Lattice QCD calculations and holographic AdS/CFT calculations.  Therefore, building the full holographic dual of the $\mathcal{N} = 1^*$ theory can provide both a novel description of many gauge-theory phenomena that had not been hitherto describable holographically, as well as a benchmark that will allow Lattice QCD to sharpen its predictions and to test the precision of the various approximations one makes in lattice calculations.

However, despite its very important physics, and despite multiple attempts over the 18 years since the pioneering work of Polchinski and Strassler \cite{Polchinski:2000uf}, the fully backreacted supergravity solution dual to any of the vacua of the $\mathcal{N} = 1^*$ theory is still not known. And there is good reason for this: the RR and NS-NS three-form flux perturbations (on top of empty $AdS_5 \times S^5$) that one has to turn on to describe just the UV of this solution break the internal isometry from $SO(6)$ to $SO(3) \times U(1)$ when the three chiral multiplet masses are equal,\footnote{When the chiral multiplet masses are not equal, the symmetry is further broken.} and furthermore, one expects from the mean-field Polchinski-Strassler analysis that the $U(1)$ will be broken in each and every supersymmetric phase of the theory. Hence, one would have to find a supersymmetric flow that is at least cohomogeneity-3, and while this has been done using supersymmetric-solution-building technology \cite{Gauntlett:2002sc,Gowdigere:2003jf} in several examples \cite{Nemeschansky:2004yh,Lin:2004nb}, the multiple field components and the complexity of the topology have so far prohibited the direct construction of the full Polchinski-Strassler solution.

It is the purpose of this paper to break this 18-year old impasse, and to construct the first fully-backreacted solution dual to one of the vacua of the $\mathcal{N} = 1^*$ theory: the {\it high-temperature deconfined vacuum}, dual to a black hole in the asymptotically $AdS_5 \times S^5$ solution with non-trivial three-form fluxes dual to the mass deformation. Finding this solution is a very non-trivial enterprise, and the most advanced work in this direction was done by Freedman and Minahan, who succeeded in computing the solution corresponding to the second-order backreaction of the three-forms on the metric, dilaton and five-form \cite{Freedman:2000xb}. The impressive analytical formulas in their work make it rather obvious that obtaining a fully-backreacted analytical solution dual to this vacuum is rather hopeless, and the absence of further progress on this problem in the last almost two decades confirms this.

Our approach to finding this solution builds on the impressive progress that has taken place over the past few years in the construction of numerical solutions.  As we will discuss in detail in Section~\ref{sec:cohomo 2}, the simplest black hole solution of the $\mathcal{N} = 1^*$ theory has cohomogeneity two, and can be described by twenty functions of two variables satisfying second-order partial differential equations. Since the solution has only a black hole and no polarized branes, its rod structure is very simple, and the domain of dependence of these functions can be mapped to the unit square. To find these functions numerically we use the DeTurck method, which consists of adding certain `DeTurck' terms to the Einstein and generalized Maxwell equation, which make them elliptic (see the reviews \cite{Headrick:2009pv,Wiseman:2011by,Dias:2015nua}). Any solution of the Einstein-Maxwell equations is also a solution of the Einstein-Maxwell-DeTurck ones, and it is easy to ascertain for which solutions the reverse holds as well.

Having obtained the full ten-dimensional solution, we can compute its free energy (and other thermodynamic quantities) and, in particular, we can investigate their behavior as one lowers the temperature and approaches the scale where one expects the confinement/deconfinement phase transition to take place.  We find evidence of a (likely first-order) phase transition, at a temperature that is about half the mass of the fermions and bosons of the $\mathcal{N} = 1^*$ theory. However, we do not know whether the phase transition we find is the deconfinement one or an intermediate one giving rise to a black hole with a different horizon topology.  The temperature of the phase transition we find is a feature that could be in principle checked by putting the theory on lattice, and hence our result constitutes the first prediction of string theory and AdS-CFT that could be independently checked by four-dimensional Lattice QCD calculations. 

Ten-dimensional (10d) type IIB supergravity compactified on a five-sphere gives rise to five-dimensional (5d) $\mathcal{N}=8$ gauged supergravity, and it is in this theory that the first holographic analysis of the $\mathcal{N} = 1^*$ theory was performed, by Giradello, Petrini, Porrati, and Zaffaroni (GPPZ) \cite{Girardello:1999bd}. Unfortunately, five-dimensional gauged supergravity cannot capture all the ten-dimensional physics. In particular, GPPZ  found a 5d \emph{singular} solution, which, as we will discuss in Section \ref{sec:vacua}, has too much symmetry to correspond to any of the supersymmetric vacua of the $\mathcal{N} = 1^*$ theory. This conclusion was confirmed in the very recent studies of \cite{Petrini:2018pjk,Bobev:2018eer} which found the full description of the 10d uplift of the GPPZ solution (thus completing the partial uplift of the GPPZ flow of \cite{Pilch:2000fu}). The 10d uplift of GPPZ solution is still singular \cite{Petrini:2018pjk,Bobev:2018eer}, and in particular, this singularity does not appear to admit an interpretation in terms of brane sources.  As such, the GPPZ flow does not describe any of the supersymmetric vacua of the $\mathcal{N} = 1^*$ theory. 

On the other hand, by analyzing the system in the full ten-dimensional context,  Polchinski and Strassler  \cite{Polchinski:2000uf} argued (in a mean-field approximation) that the singularity of the GPPZ solution should be resolved by the Myers-effect polarization \cite{Myers:1999ps} of the D3 branes underlying the $AdS_5 \times S^5$ geometry into five-branes \cite{Polchinski:2000uf}.  Refs. \cite{Petrini:2018pjk,Bobev:2018eer} show that the supersymmetric vacua proposed by Polchinski and Strassler do not fit within the subsector of solutions resulting from the uplift of GPPZ, and Ref. \cite{Petrini:2018pjk} in addition considers a larger, four-scalar truncation (which allows the GPPZ scalars to be complex), leading again to the same negative result.  The maximal consistent truncation with the same $SO(3)$ symmetry as GPPZ contains eight scalars, and while it has not been conclusively shown that this truncation does \emph{not} contain the smoothly-resolved polarized branes of Polchinski-Strassler, the evidence for such a conclusion is fairly strong.  Thus finding the holographic description of the  supersymmetric vacua of $\mathcal{N} = 1^*$ proposed in \cite{Polchinski:2000uf} will require directly solving the IIB equations in ten dimensions without imposing truncations of the KK modes.  
  
Therefore it is established that five-dimensional gauged supergravity has not been (and probably will not be) able to capture the physics of $\mathcal{N} = 1^*$ supersymmetric vacua. However, in this manuscript, we will show that it \emph{can} very well capture the physics of the high-temperature, deconfined phase vacuum (or at least of one such deconfined vacuum). 

Constructing the black hole in the GPPZ truncation of five-dimensional $\cN=8$ gauged supergravity is an order of magnitude easier than constructing the full black hole solution in ten dimensions. Indeed, the five-dimensional black hole solution is described by four functions of one variable, which are much easier to find numerically than the 20 functions of two variables describing the ten-dimensional solution.  In view of this fact, after reviewing the $\cN = 1^*$ super-Yang-Mills theory in Section \ref{sec:review}, we will first describe (in Section \ref{sec:5d}) the solutions constructed in five-dimensional supergravity, leaving the construction of the full ten-dimensional solution to Section \ref{sec:IIBflows}. We will then confirm that the thermodynamics of the five- and ten-dimensional solutions agree.  Alternatively, one could arrive at the same conclusion by applying the uplift formulas of \cite{Baguet:2015sma}, as was done in \cite{Petrini:2018pjk,Bobev:2018eer} (which were published while this work was being completed), to obtain the ten-dimensional solution for the high-temperature deconfined vacuum (of Section \ref{sec:IIBflows}) from its five-dimensional counterpart of Section \ref{sec:5d}.  However, our brute-force approach of directly solving the IIB equations of motion in ten dimensions sets the stage for further searches for solutions which do not have a dimensionally-reduced description in five-dimensional gauged supergravity.
 
%%%%%%%%%%%%%%
\section{$\cN = 1^*$ super-Yang-Mills and its holographic dual}
\label{sec:review}
%%%%%%%%%%%%%%

The conformal, $\cN=4$ super-Yang-Mills theory in 4 dimensions with gauge group $SU(N)$ is holographically dual, at large $N$ and large 't Hooft parameter, to type IIB string theory in $AdS_5 \times S^5$.  Deformations of the $\cN=4$ theory by relevant operators correspond on the gravity side to deformations of $AdS_5 \times S^5$ by non-normalizable modes. We are interested in a deformation that preserves at least $\cN=1$ supersymmetry, and corresponds to adding to the superpotential ${\cal W}$ the contribution
\begin{equation} \label{W deformation}
\Delta {\cal W} =m_1 \tr \Phi_1^2 + m_2 \tr \Phi_2^2 + m_3 \tr \Phi_3^2\,.
%\Delta {\cal W} = \frac{1}{g_{YM}^2} (m_1 \tr \Phi_1^2 + m_2 \tr \Phi_2^2 + m_3 \tr \Phi_3^2)\,.
\end{equation} 
In $\cN = 1$ language this deformation gives masses $m_1, m_2, m_3$ to the three chiral multiplets. When $m_1 = m_2 \neq 0$ and $m_3 = 0$ the supersymmetry is enhanced to $\cN=2$, and the corresponding flow is known as the Pilch-Warner flow \cite{Pilch:2000ue}. When only one chiral multiplet mass is non-zero, the theory flows in the infrared to an $SU(2) \times U(1)$-invariant (Leigh-Strassler) conformal fixed point \cite{Leigh:1995ep}.  Generically, when all three $m_i \neq 0$, the theory is called the $\cN=1^*$ theory, and it has a rich (and well-studied) structure of vacua.

%%%%%%%%%%%%%%
\subsection{The vacua of the $\cN=1^*$ theory and their supergravity duals \label{sec:vacua}}
%%%%%%%%%%%%%%

Assuming all the masses in \eqref{W deformation} are nonzero, one can rescale the fields to set them equal, and the classical vacua are described by solutions to the equations
\begin{equation} \label{vacuum eqs}
[\Phi_i, \Phi_j] = - \frac{m}{\sqrt{2}} \varepsilon_{ijk} \Phi_k \,.
\end{equation}
These are equivalent to the commutation relations of $SU(2)$, and thus the (classical) vacua are classified by homomorphisms of $SU(2)$ into the gauge group $SU(N)$. 
These have a nice combinatoric structure, which we will not detail here (but we refer the reader to \cite{vafa1994strong}).  In particular, there exist special vacua for each positive integer $d$ that divides $N$, which have (classically) an unbroken $SU(d)$ gauge symmetry.  Quantum mechanically, these vacua split into $d$ separate vacua with totally broken gauge symmetry and a mass gap \cite{Donagi:1995cf}.  They can be described as \emph{Higgs} or \emph{screening} ($d = 1$), \emph{confining} ($d = N$), or \emph{oblique confining} ($1 < d < N$) vacua, wherein various combinations of electric/magnetic charges are confined/screened.

Polchinski and Strassler argued \cite{Polchinski:2000uf}  that in the bulk, the three families of vacua of the $\cN=1^*$  theory should correspond to brane polarization of the D3 branes underlying the  $AdS_5 \times S^5$ geometry into spherical shells  with D5, NS5 or more-general  $(p,q)$ 5-brane dipole charges, respectively. This polarization is caused by the RR and NS-NS three-form non-normalizable modes dual to the deformation \eqref{W deformation}. The Higgs vacua in which the D3 branes are polarized into D5 branes wrapping an $S^2$ equator of the $S^5$ can also be described from a field theory perspective  as solutions to the equations \eqref{vacuum eqs}. However, the confining vacua (corresponding to NS5-branes wrapping the orthogonal $S^2$ equator) and the oblique vacua (corresponding to 
$(p,q)$ 5-branes wrapping an $S^2$ equator of oblique orientation) are `quantum vacua' \cite{Donagi:1995cf} which cannot be described as classical solutions of \eqref{vacuum eqs}.

The Polchinski-Strassler analysis was triggered by an earlier attempt by Girardello, Petrini, Porrati, and Zaffaroni (GPPZ)  \cite{Girardello:1999bd} to describe the $\cN=1^*$  theory using five-dimensional gauged supergravity. The most generic $\cN=8 $ five-dimensional gauged supergravity comes from reducing type IIB supergravity on $S^5$ by means of a consistent truncation of the Kaluza-Klein modes. Setting the three masses $m_i$ of \eqref{W deformation} equal to each other gives the problem an extra $SO(3)$ symmetry\footnote{This $SO(3)$ is a subgroup of the original $SO(6)$ $R$-symmetry.} and gives rise to a smaller truncation, which only contains two scalars\footnote{Holographically, the asymptotic behaviors of these two scalars encode the mass deformation and the vev of the gaugino bilinear:
\begin{equation}
\mu = m_1 = m_2 = m_3, \qquad \sigma \sim \langle \lambda^4 \lambda^4 \rangle.
\end{equation}}, $\mu$ and $\sigma$. GPPZ then found a surprisingly simple solution of the BPS flow equations for $\mu$ and $\sigma$ which is, however, \emph{singular} in the infrared. 

The precise relationship between the five-dimensional supergravity solution GPPZ~\cite{Girardello:1999bd} and the ten-dimensional brane configurations dual to the screening, confining and oblique vacua proposed in \cite{Polchinski:2000uf} was, until quite recently, somewhat obscure. Since the five-dimensional gauged supergravity used in \cite{Girardello:1999bd} is a consistent truncation of type IIB supergravity on $S^5$, one should in principle be able to uplift this solution into 10 dimensions and study its properties there. A \emph{partial} uplift of the solution, that gives only the ten-dimensional metric and axion-dilaton but not the IIB $p$-forms, has been known for a while  \cite{Pilch:2000fu} (see also \cite{Baguet:2015sma}). This partial uplift  does not appear to admit an interpretation in terms of the polarized branes described in \cite{Polchinski:2000uf}. The main reason is that the polarized branes of \cite{Polchinski:2000uf} appear to source normalizable modes in the {\bf 20} of SO(6) (corresponding to traceless scalar bilinears), which are absent in the GPPZ truncation. The uplift of the GPPZ solution could therefore only correspond to a smearing of the polarized branes. Since the charges of the dipole branes are quantized, the orientations of the $S^2$ equators inside the $S^5$ are rational numbers \cite{Polchinski:2000uf}, and hence the supersymmetric solution with smeared branes does not correspond to any supersymmetric vacuum of the $\cN=1^*$ theory. 

Some very recent work has shed some light on this situation: the \emph{full} uplift of the GPPZ flow solutions was finally found in \cite{Petrini:2018pjk,Bobev:2018eer}, including the IIB $p$-forms.  These results contradict some of the detailed findings of \cite{Pilch:2000fu} regarding the singularity, but they do confirm the main lesson learned---namely, that the uplifted GPPZ singularity does not admit a description in terms of anything resembling the Polchinski-Strassler polarized branes.  The authors of \cite{Petrini:2018pjk} further investigate a slight generalization of GPPZ which allows the two scalars $\mu$ and $\sigma$ to be complex, but this does not alter the conclusions appreciably.  A fully general analysis should include all 8 scalars of the $SO(3)$-invariant truncation (as will be discussed in Sec.~\ref{sec:5d sugra truncations}).  However, we suspect that these other scalars will correspond in ten dimensions to shape modes that merely move the singularities around, and turning them on will still not be sufficient to reproduce the singularity-resolution mechanism envisioned by Polchinski and Strassler.  Thus to obtain the holographic description of supersymmetric $\mathcal{N} = 1^*$ vacua proposed in \cite{Polchinski:2000uf} will most likely require directly solving the IIB equations of motion in ten dimensions without truncating the KK modes.

Despite this inability of five-dimensional gauged supergravity to capture the (non-singular) supersymmetric physics of the $\cN = 1^*$ theory, we will show that the high-temperature phase of this theory, which is holographically dual to a solution with a black hole, can in fact be described correctly in  five-dimensional gauged supergravity. From the perspective of a five-dimensional gauged supergravity aficionado, this is one of the main results of our paper.  We will not use the 5d-to-10d uplift machinery of  \cite{Pilch:2000fu,Petrini:2018pjk,Bobev:2018eer} directly, as it is rather complicated (and was still not available while this work was being completed).  We will rather construct the simplest solutions (with nontrivial $m$ and $\sigma = 0$) both in five and ten dimensions, and show that their physics is the same (in five dimensions we also construct the solutions with $m\neq 0$ and $\sigma \neq 0$). We will then argue that five-dimensional gauged supergravity should correctly describe the high-temperature physics for more generic deformations, construct the corresponding five-dimensional black hole solutions, and analyze their physics. In the remainder of this section we will  summarize key aspects of the five- and ten-dimensional theories, as well as the GPPZ solution and its IIB uplift.
 
%%%%%%%%%%%%%%
\subsection{Five-dimensional  gauged supergravity and GPPZ flow}
\label{sec:5d sugra truncations}
%%%%%%%%%%%%%%

The maximal 5d $\cN = 8$ supergravity was first constructed in \cite{Gunaydin:1984qu,Pernici:1985ju,Gunaydin:1985cu}, and was shown to be obtained from type IIB supergravity via Kaluza-Klein reduction on $S^5$.  Such reductions (on curved manifolds) are not generically possible---there is an infinite tower of massive KK modes from the perspective of the lower-dimensional theory---and in order to have a theory with a finite number of fields, one needs a `consistent truncation' of this tower to exist.  This only happens in special circumstances, such as when the internal manifold is a Lie group manifold, or in the famous near-horizon geometries $AdS_5 \times S^5$, $AdS_4 \times S^7$ and $AdS_7 \times S^4$ (see, for example, \cite{Cvetic:2000dm,Gibbons:2003gp,deWit:2013ija,Godazgar:2013pfa,Godazgar:2013nma,Godazgar:2013dma,Godazgar:2013oba,Lee:2014mla}).

The bosonic content of $\cN=8$ supergravity in 5 dimensions consists of the metric, 12 two-form fields, 15 gauge fields, and 42 scalars parameterizing the coset $E_{6(6)}/USp(8)$.  Ignoring the gauge fields and tensors, the scalar subsector of the theory has the Lagrangian\footnote{Note that we use `mostly plus' conventions throughout this work.}
\begin{equation}
\cL = \sqrt{-g} \Big( R - \frac{1}{24} \nabla_\mu \cM_{MN} \nabla^\mu \cM^{MN} - V(\cM) \Big),
\end{equation}
where $\cM_{MN}$ is a $27 \times 27$ matrix of $E_{6(6)}$ parameterized by the scalars, and $V$ is a potential which is some polynomial of $\cM_{MN}$ and its inverse.  The scalars are related to modes of the $S^5$ metric, of the axion-dilaton, as well as of other type IIB form fields that have no legs along the 5d `external' directions, although the precise relationship is highly nontrivial.

The first step is to impose some symmetry to reduce the number of scalars.  The $\cN=4$ super-Yang-Mills theory has an $SU(4) \simeq SO(6)$ $R$ symmetry which is generically broken by turning on masses for the chiral multiplets as in \eqref{W deformation}.  However, for the choice of \emph{equal} masses, we can preserve an $SO(3)$ symmetry.\footnote{This is of course no longer an $R$ symmetry, but simply a global symmetry that rotates the massive chiral multiplets among each other.}  The particular $SO(3)$ is the one which is `diagonal' in $SO(6)$; or equivalently, it is the real subgroup $SO(3) \subset SU(3) \subset SO(6)$, under which the \textbf{4} of $SU(4)$ decomposes as $\mathbf{4} \to \mathbf{3} + \mathbf{1}$ and the \textbf{6} decomposes as $\mathbf{6} \to \mathbf{3} + \mathbf{3}$.  Its action on the super-Yang-Mills scalars $X^I$ is easiest to describe in terms of the complex combinations
\begin{equation}
Z^1 \equiv \frac{X^1 + i X^2}{\sqrt2}, \qquad Z^2 \equiv \frac{X^3 + i X^4}{\sqrt2}, \qquad Z^3 \equiv \frac{X^5 + i X^6}{\sqrt2};
\end{equation}
then the $Z^i$ rotate as a vector of $SO(3)$.  As explained in \cite{Pilch:2000fu}, after imposing this $SO(3)$ symmetry, the 42 scalars of the 5d theory are reduced to 8, which correspond to the following operators in the holographic $\cN=1^*$ super-Yang-Mills:  the gauge coupling, the instanton angle, the two real scalar bilinears
\begin{align} \label{scalar bilinears}
\cO_1 &= \sum_{i=1}^3 \big( \tr(X^{2i-1} X^{2i-1}) - \tr(X^{2i} X^{2i}) \big) = \sum_{i=1}^3 \tr \big( (Z^i)^2 + (\bar Z^i)^2 \big), \\
\cO_2 &= 2 \sum_{i=1}^3 \tr(X^{2i-1} X^{2i}) = -i \sum_{i=1}^3 \tr \big( (Z^i)^2 - (\bar Z^i)^2 \big);
\end{align}
and the two complex fermion bilinears,
\begin{equation} \label{fermion bilinears}
\cO_3 = \sum_{i=1}^3 \tr(\lambda^i \lambda^i), \quad \text{and} \quad \cO_4 = \tr(\lambda^4 \lambda^4),
\end{equation}
where $\lambda^i$, for $i=1,2,3$, are chiral superfields, and $\lambda^4$ is the gaugino.  We note that there is also a $U(1)$ part of the $SO(6)$ which commutes with the $SO(3)$; it rotates the $Z^i \to e^{i \alpha} Z^i$ by a global phase, and transforms these operators as follows:\footnote{Although it is not obvious from the field theory Lagrangian, this $U(1)$ also acts nontrivially on the gauge coupling and instanton angle, via the `bonus' $U(1)_Y \subset SL(2,\RR)$ of the enhanced $S$-duality group at large $N$ \cite{Intriligator:1998ig}.  This follows from the flow equations.} 
\begin{equation} \label{u1 operators}
\cO_1 + i \cO_2 \to e^{2i \alpha} (\cO_1 + i \cO_2), \qquad \cO_3 \to \cO_3, \qquad \cO_4 \to e^{3i \alpha} \cO_4.
\end{equation}
Thus, the simplest non-trivial truncation is the one that imposes $SO(3) \times U(1)$ symmetry, leaving only the operator $\cO_3$, which corresponds to the mass of the chiral multiplets.

GPPZ considered a further truncation of these eight scalars down to two: it consists of $\cO_3$ and $\cO_4$, denoted respectively by $\mu$ and $\sigma$ (which can be chosen to be real), and is obtained by imposing a \emph{discrete} symmetry on the above set of eight $SO(3)$-invariant scalars.  The Lagrangian of this truncated theory is 
\begin{equation}
\label{eq:5daction}
S = \frac{1}{2\kappa_5^2} \int d^5x \sqrt{-g} \Big[ R - 2(\nabla \mu)^2 - 2(\nabla \sigma)^2 - 4 V \Big] ,
\end{equation}
where $\kappa_5^2 =8\pi G_5= \kappa_{10}^2/\text{vol}(S^5) = \kappa_{10}^2/\pi^3$ (with $G_5$ being Newton's constant in $d=5$) and the potential is
\begin{equation}
\label{eq:GPPZpotential}
V =-\frac{3}{8} \bigg[4 \cosh (2 \sigma ) \cosh \bigg(\frac{2 \mu }{\sqrt{3}}\bigg)-\cosh ^2(2 \sigma )+\cosh ^2\left(\frac{2 \mu }{\sqrt{3}}\right)+4\bigg].
\end{equation}
To obtain \emph{supersymmetric} solutions, GPPZ rewrote this potential in terms of a superpotential, ${\cal W}$ and, via the appropriate manipulations, arrived at a system of first-order ODEs which could be integrated.  They found two solutions, the first one with $\sigma = 0$, and the second one with $\sigma \neq 0$.  The metric for both solutions is
\begin{equation}
\dd s^2 = \dd \tilde y^2 + e^{2\tilde \phi(\tilde y)} \eta_{\mu\nu} \dd x^\mu \dd x^\nu, \qquad \mu, \nu = 0,1,2,3.
\end{equation}
The solution with $\sigma=0$ is
\begin{subequations}\label{soln:GPPZ}
	\label{eq:super1}
	\begin{align}
	&\tilde{\phi}(\tilde{y})=\frac{1}{2} \big( y+\log [2\sinh(\tilde{y}-C_1)] \big),
	\\
	& \mu(\tilde{y}) = \frac{\sqrt{3}}{2}\log \bigg[\frac{1+e^{-(\tilde{y}-C_1)}}{1-e^{-(\tilde{y}-C_1)}} \bigg],
	\\
	& \sigma(\tilde{y}) =0,
	\end{align}
\end{subequations}
and the solution with nontrivial  $\sigma$ is
\begin{subequations}\label{soln:GPPZgaugino}
	\label{eq:super2}
	\begin{align}
	&\tilde{\phi}(\tilde{y})=\frac{1}{2}\log [2\sinh(\tilde{y}-C_1)]+\frac{1}{6} \log [2\sinh(3\tilde{y}-C_2)],
	\\
	& \mu(\tilde{y}) = \frac{\sqrt{3}}{2}\log \bigg[\frac{1+e^{-(\tilde{y}-C_1)}}{1-e^{-(\tilde{y}-C_1)}} \bigg],
	\\
	& \sigma(\tilde{y}) = \frac{1}{2}\log \bigg[\frac{1+e^{-(3\tilde{y}-C_2)}}{1-e^{-(3\tilde{y}-C_2)}} \bigg],
	\end{align}
\end{subequations}
where $C_1$ and $C_2$ are integration constants.\footnote{The first solution can be recovered from the second by shifting $\tilde \phi(\tilde y)$ by a constant (which can then be absorbed by rescaling the coordinates $x^\mu$), and then taking $C_2 \to - \infty$.}  In the UV, the functions $\mu$ and $\sigma$ have the expected decay to encode the mass deformation and a gaugino VEV:
\begin{equation}
\mu \sim e^{-\tilde y}, \qquad \sigma \sim e^{-3 \tilde y} \quad \text{as} \quad \tilde y \to \infty.
\end{equation}
However, in the IR the metric becomes singular. Furthermore note that the GPPZ solutions \eqref{soln:GPPZ} and \eqref{soln:GPPZgaugino} have a scalar $\mu$ with a source but the VEV of the dual operator vanishes. This will be important when analyzing the results for the VEV  of $\mathcal{O}_3$ plotted in Figs.~\ref{fig:thermo1} and \ref{fig:thermo5b}. In addition, it is useful to define the following dimensionless quantity
\begin{equation}
\lambda \equiv \frac{\pi^2}{3\sqrt{3}m^3}\langle \mathcal{O}_4\rangle\,,
\label{eq:lambda}
\end{equation}
which was introduced in \cite{Pilch:2000fu} as tool to understand the singularity behavior of (\ref{soln:GPPZgaugino}) from a ten-dimensional perspective. In fact, the authors of \cite{Girardello:1999bd} argue that good flows must have $\lambda\leq1$. We shall later see that our finite-temperature solutions approach $\lambda = 1$ from below in a particular singular limit.

%%%%%%%%%%%%%%
\subsection{The uplift of the supersymmetric GPPZ solution to ten dimensions}
%%%%%%%%%%%%%%

%\oscar{XXX I did minor changes in this Sec. to accommodate \cite{Petrini:2018pjk,Bobev:2018eer}. Please check XXX}

Applying the uplift \emph{ans\"atze} of \cite{Gunaydin:1985cu}, Pilch and Warner obtained the 10-dimensional metric and axion-dilaton corresponding to the GPPZ flows \cite{Pilch:2000fu}. However, the IIB $p$-forms describing the uplifted GPPZ solution were only obtained recently (in parallel) by Petrini, Samtleben, Schmidt and Skenderis \cite{Petrini:2018pjk} and by one of the authors, Bobev, Gautason and van Muiden \cite{Bobev:2018eer}.  Here we present only a sketch of the metric and axion-dilaton matrix, and leave the considerably more complicated details to those references.

The 10d GPPZ Einstein-frame metric takes the form
\begin{equation} \label{uplift metric}
\dd s_{10}^2 = \xi^{1/2} \left(e^{2A(r)} \eta_{\mu\nu} \dd x^\mu \dd x^\nu + \dd r^2\right) + \xi^{-3/2} \, \dd \widetilde \Omega_5^2,
\end{equation}
where the term in parentheses is the 5d GPPZ metric (with $r = \tilde y$ and $A = \tilde \phi$), $\dd \widetilde \Omega_5^2$ is a deformed metric on $S^5$ (which depends on $r$ as well as its own internal coordinates), and $\xi$ is a function of both $r$ and the internal coordinates.  The deformed $S^5$ metric can be written as\footnote{This formula corrects a minor typo in the second term on the second line of (6.2) of \cite{Pilch:2000fu}.}
\begin{equation} \label{PW metric}
\begin{split}
\dd \widetilde \Omega_5^2 &= a_1 \, \dd u^i \dd u^i + 2 a_2 \, \dd u^i \dd v^i + a_3 \, \dd v^i \dd v^i \\
& \qquad + a_4 \, \big( \dd (\vec u \cdot \vec v) \big)^2 + 2a_5 \, (v^i \dd u^i) (u^j \dd v^j) + 2a_6 \, (u^i \dd u^i)(v^j \dd v^j),
\end{split}
\end{equation}
where $u^i, v^i$ are two vectors in $\RR^3$ constrained such that $u^2 + v^2 = 1$ (thus describing a unit $S^5$ in $\RR^6 \simeq \RR^3 \times \RR^3$), and the functions $a_1$ through $a_6$, as well as the warp factor $\xi$, are complicated, but known, functions of the radial coordinate $r$ and the $SO(3)$ invariants $u^2 - v^2$ and $\vec u \cdot \vec v$.  The axion-dilaton matrix is given in terms of the same functions,
\begin{equation}
\cM = \frac{1}{\xi}
\begin{pmatrix}
a_1 & a_2 \\ a_2 & a_3
\end{pmatrix}.
\end{equation}
Written in this way, the metric and axion-dilaton are manifestly $SO(3)$-invariant.
We refer the reader to \cite{Petrini:2018pjk,Bobev:2018eer} for a discussion of the IIB $p$-forms that describe the uplifted GPPZ solution.  

In the limit $\sigma \rightarrow 0$, one can show that the metric simplifies to a form which is (additionally) invariant under $SO(2)$ rotations between $\vec u$ and $\vec v$, and thus the isometry group enhances to $SO(3) \times U(1)$, matching expectations from the consistent truncation.  The axion-dilaton matrix $\cM$, however, changes by an $SO(2)$ rotation under the $U(1)$ part of this isometry action.  In Polchinski-Strassler language, this corresponds to the rotation of D5-branes into NS5-branes, or of the Higgs vacuum into the confining vacuum, moving through the intermediate $(p,q)$ oblique vacua.  Therefore the $\sigma = 0$ solutions which are \emph{invariant} under this $U(1)$ should somehow correspond to an unusual distribution of $(p,q)$ brane charges which vary continuously over the circle $p^2 + q^2 = N^2$ as one moves around a circle in the geometry.  As we will explain below, such continuous distributions of charges in $(p,q)$ space are not possible when $N$ is finite.\footnote{The analysis in \cite{Pilch:2000fu} of the IR singularity of the metric in \eqref{uplift metric} and \eqref{PW metric} indeed finds such a distribution of $(p,q)$ charges, but the behavior of the metric is not consistent with the 5-branes expected from Polchinski-Strassler.}

%%%%%%%%%%%%%%
\subsection{The 5D-10D connection}
%%%%%%%%%%%%%%

As described in \cite{Pilch:2000fu}, the IR limit of the metric \eqref{uplift metric} contains singularities which cannot be easily interpreted as coming from the polarized Polchinski-Strassler branes.  While the analysis of \cite{Petrini:2018pjk,Bobev:2018eer} disagrees with \cite{Pilch:2000fu} on some details, the same conclusion is recovered: even after uplifting, GPPZ contains singularities which are simply incompatible with the Polchinski-Strassler construction.  There are two ways to see this: first, the symmetry of the GPPZ solution is only realized in ten dimensions if the five-branes are smeared on $S^2$ equators of the $S^5$ while being transformed by $SL(2,R)$ S-duality group. Since only an  $SL(2,Z)$ subgroup is consistent with quantized five-brane charges, the smeared configuration must necessarily have non-quantized charges.  The second is to examine the fields sourced by say D5 branes wrapping an $S^2$ equators of the $S^5$. These fields must contain metric and dilaton modes transforming in the {\bf 20} of $SO(6)$, and hence the $SO(3)$-invariant $L=2$ scalar modes corresponding to $\cO_1$ and $\cO_2$ should be also turned on. 

It is possible that when including these extra modes certain ten-dimensional solutions would be indeed captured by five-dimensional supergravity, which would be quite remarkable.  However, in the absence of an explicit uplift of these modes, it is difficult to ascertain whether this is true.  It may also be that five-dimensional supergravity cannot describe ten-dimensional configurations with polarized branes. We leave the exploration of this important issue to future work, and rather focus on a phase of the $\cN=1^*$ theory whose enhanced symmetry should make it more amenable to a five-dimensional description: the high-temperature phase. 

One expects this phase to be described in the bulk by a black hole with $\mathbb{R}^3 \times S^5$ horizon cross-sectional topology, which dominates the grand canonical ensemble.  Below some critical temperature these black holes will no longer dominate (although they may still exist as solutions), and for lower temperatures one may hope to find black objects that could come from the blackening of the polarized branes of the zero-temperature phase, which would be black ringoids with nontrivial five-brane dipole charge and $\mathbb{R}^3 \times S^2 \times S^3$ horizon topology. Since the black hole with $\mathbb{R}^3 \times S^5$ horizon topology is very symmetric and sources no $L=2$ mode, one may expect that its physics is correctly captured by the GPPZ five-dimensional supergravity truncation, and we will show that this is indeed so by explicitly constructing both the ten-dimensional solution and the five-dimensional one, and giving several arguments they are the same. It is again unclear whether the the black ringoids with less symmetries would be describable in five dimensions, but, even if they are not,  it may be possible to get a glimpse of the phase where they dominate by looking at perturbations of the spherical black holes.

In the following sections we will construct finite-temperature GPPZ flows, as well as type IIB black holes in the $\sigma = 0$ limit with enhanced $SO(3) \times U(1)$ symmetry. Then we will also construct black holes solutions that have $\sigma \neq 0$ (in addition to finite $\mu$). We will show that the simplest GPPZ model indeed captures all of the relevant physics at sufficiently-high temperature.

%%%%%%%%%%%%%%
\section{The $SO(3)$-invariant flows at finite temperature in 5d}
\label{sec:5d}
%%%%%%%%%%%%%%

In this section we use numerical methods to study flows in the $SO(3)$-invariant subsector of the 5d supergravity theory at \emph{finite} temperature. We might say that our thermal phases describe finite-temperature phases of ${\cal N}=1^*$ super-Yang-Mills theory in the GPPZ subsector. In the sense that our solutions only have the scalars  $\mu$ and (eventually) $\sigma$ turned on, corresponding to having the operators $\cO_3$ and $\cO_4$ of \eqref{fermion bilinears} switched on. There might well exist thermal phases with flows that have more  $SO(3)$-invariant scalars turned on (that would complete the phase diagram of thermal solutions of the theory with this symmetry) but we leave the exploration of these phases for future work.

%%%%%%%%%%%%%%
\subsection{The 2-scalar GPPZ subsector}
%%%%%%%%%%%%%%

The GPPZ subsector of the theory has only $\mu$ and $\sigma$ turned on, dual to the operators $\cO_3$ and $\cO_4$ of \eqref{fermion bilinears}.  The action is given in \eqref{eq:5daction}, and the equations of motion obtained from this action are
\begin{subequations}
	\label{eq:EOMs}
	\begin{align}
	&R_{ab}=2\nabla_a \mu\nabla_b \mu + 2\nabla_a \sigma\nabla_b \sigma+\frac{4}{3} V g_{ab},
	\\
	& \nabla^2 \mu = \frac{\partial V}{\partial \mu},
	\\
	& \nabla^2 \sigma = \frac{\partial V}{\partial \sigma},
	\end{align}
\end{subequations}
where the potential $V$ was given in \eqref{eq:GPPZpotential}. Expanding this potential to quadratic order in the fields yields
\begin{equation}
V \approx -3 -\frac{3}{2} \left(\mu^2+\sigma ^2\right)+\ldots
\end{equation}
The constant term informs us that the theory admits aysmptotically-$AdS_5$ solutions and, in particular, that their radius is $L=1$. On the other hand, the quadratic term tells us that $\mu$ and  $\sigma$ each have mass $M^2 = -3$.  The boundary expansion for a scalar of mass $M$ in a unit-radius, asymptotically-$AdS_5$ spacetime in the Fefferman-Graham coordinate $z$ is
\begin{equation}
\phi(z, t, x) = z^{\Delta_+} \left(a_0(x,t) + a_1(x,t) z + ... \right) + z^{\Delta_-}   \left(b_0(x,t) + b_1(x,t) z + ... \right), 
\end{equation}
where the conformal dimensions are given by
\begin{equation}
\Delta_{\pm} = 2 \pm \sqrt{4 + M^2}\,.
\end{equation}
Thus, for both $\mu$ and $\sigma$, one has $\Delta_+ = 3$ and $\Delta_- = 1$.  The $\Delta_+$ branch is normalizable and the $\Delta_-$ one is non-normalizable. Thus, from the usual AdS/CFT prescription, we conclude that $a_0$ describes a VEV, and $b_0$ describes a source.

The scalar $\mu$ corresponds to the equal-mass deformation of the field theory \eqref{W deformation}, so we expect it to have a source.  However, adding a source for $\sigma$ (and thus a mass to the gaugino) would \emph{explicitly} break supersymmetry of the $\cN=1^*$ theory to $\cN=0^*$, which we do not want to do.  Therefore $\sigma$ should have only a VEV.  Thus we expect the following asymptotic behavior near the boundary:
\begin{equation}
\mu \sim m \,z, \qquad \sigma \sim z^3,
\end{equation}
where $m = m_1 = m_2 = m_3$ will be the mass of the chiral multiplets.

%%%%%%%%%%%%
\subsection{Constructing finite-temperature flows in the 2-scalar GPPZ subsector}
\label{sec:5d setup}
%%%%%%%%%%%%

First let us discuss \emph{ans\"atze} and gauge fixing.  We will construct solutions to (\ref{eq:EOMs}) with planar symmetry, whose metric has the general form
\begin{equation} \label{eq:ansatz5D}
\mathrm{d}s^2 = \frac{1}{y}\left[-f(y)\mathrm{d}t^2+\frac{\mathrm{d}y^2}{4\,y\,g(y)}+R(y)^2\Big(\mathrm{d}w_1^2+\mathrm{d}w_2^2+\mathrm{d}w_3^2\Big)\right]
\end{equation}
and $\mu$ and $\sigma$ depend on $y$ only. So far, we have not fixed our gauge freedom, since the form of line element above is completely invariant under reparameterizations of $y$. We choose $R(y)=1$, the so-called Schwarzschild gauge. Indeed, the familiar AdS$_5$ planar black hole can be written in these coordinates if we take  
\begin{equation}\label{eq:5dplanarBH}
f(y)=g(y)=1-\frac{y^2}{y_0^2}
\end{equation}
with the black hole horizon being the null hypersurface $y=y_0$. The metric \emph{ansatz} (\ref{eq:ansatz5D}) with $R=1$ still enjoys a scaling symmetry of the form $y\to \lambda y$, $w_i\to\lambda w_i$ and $t\to\lambda t$, for constant $\lambda\in\mathbb{R}$. We use this freedom to fix the horizon at $y_0=1$.

It is useful to eliminate the hyperbolic functions $\mu(y)$ and $\sigma(y)$ in favor of algebraic ones, $q_2(y)$ and $q_3(y)$, through the redefinitions
\begin{equation}\label{mu-q2}
\mu \equiv \sqrt{3}\,\mathrm{arcsinh}\left[\frac{\sqrt{y}\,q_2(y)}{\sqrt{3}}\right],\qquad \text{and}\qquad\sigma \equiv \mathrm{arcsinh}\left[y^{3/2}q_3(y)\right].
\end{equation}
To ensure we have a horizon at $y=1$, we further define
\begin{equation}
f(y)\equiv (1-y^2)q_1(y),\qquad \text{and}\qquad g(y)\equiv (1-y^2)q_4(y),
\end{equation}
where $q_1$ and $q_4$ are smooth functions at $y=1$. Thus we have to solve for four unknown functions of $y$,  $\{q_1,q_2,q_3,q_4\}$. Our boundary conditions at the conformal boundary, located at $y=0$, demand
\begin{equation}
q_1=q_4=1,\qquad \text{and}\qquad q_2 = m.
\end{equation}
As for $q_3$, our \emph{ansatz} is written in such a way that $y \to z^2$ asymptotically, and thus $\sigma \sim z^3$ alread has the correct fall-off.  Expanding the equations of motion close to $y=0$ reveals that the boundary condition for $q_3$ is
\begin{equation}
\left.\partial_y q_3 +\frac{m^2}{2}q_3\right|_{y=0}=0.
\end{equation}

At the horizon we will impose regularity by demanding that all $q_i$ have a regular Taylor expansion around $y=1$.  Plugging the above \emph{ans\"atze} into the Einstein-scalar equations of motion \eqref{eq:EOMs} gives two first-order differential equations for $q_1$ and $q_4$ and two second-order differential equations for $q_2$ and $q_3$. We use the first-order ODE for $q_1$ to express $q_4$ as an algebraic function of $q_1$, $q_2$, $q_3$ and their first derivatives. We are thus left with a system of three coupled second-order nonlinear differential equations in $y$. At the horizon we obtain Robin boundary conditions for $q_1$, $q_2$ and $q_3$.

%%%%%%%%%%%%%%%%%%%%%%%%%%%%%%%%%%%%%
\subsubsection{Asymptotic charges and expectation values}
%%%%%%%%%%%%%%%%%%%%%%%%%%%%%%%%%%%%%

Next we want to know how to extract physical quantities from our solutions.  To obtain the asymptotic quantities, we start by solving the second-order equations of motion for $q_1$, $q_2$ and $q_3$ close to $y=0$ as a power series in $y$ of the form:
\begin{equation}
q_1 = \sum_{j=0}^{+\infty}y^j\,q_1^{(j)},\quad q_2 = \sum_{j=0}^{+\infty}y^j\,q_2^{(j)}, \quad\text{and}\quad q_3 = \sum_{j=0}^{+\infty}y^j\,q_3^{(j)}.
\end{equation}
Order by order in $y$, we find
\begin{subequations}
	\begin{align}
	&q_1(y) = 1+ q_1^{(2)}\,y^2+O\left(y^3\right),
	\\
	&q_2(y)=m+q_2^{(1)}\,y+\frac{m}{24} \left(3-2 m q_2^{(1)}-3 q_1^{(2)}\right)\,y^2 +O\left(y^3\right),
	\\
	&q_3(y)=q_3^{(0)}-\frac{1}{2} m^2 q_3^{(0)}\,y+\frac{q_3^{(0)} }{24} \left(9+5m^4-14 m q_2^{(1)}-9 q_1^{(2)}\right)\,y^2 +O\left(y^3\right),
	\end{align}
\end{subequations}
where $q_1^{(2)}$, $q_2^{(1)}$ and $q_3^{(0)}$ are free parameters. All higher-order terms can be shown to be algebraic functions of $q_1^{(2)}$, $q_2^{(1)}$, $q_3^{(0)}$ and $m$. Note that  since the boundary conditions at the horizon are of Robin type, we have three free constants there, which we can take to be $q_1(1)$, $q_2(1)$ and $q_3(1)$. Thus, for fixed $m$ we expect a unique solution.

We change to Fefferman-Graham coordinates by introducing an expansion of $y$ in terms of a new variable, $z$:
\begin{equation}
y=z^2 \sum_{j=0}^{+\infty}\alpha_j z^j,
\end{equation}
and demanding that $g_{zz}=1/z^2$ to all orders in $z$. This gives
\begin{equation}
y=z^2+\frac{m^2}{3} z^4+\frac{1}{36} \left(18 m q_2^{(1)}+9 q_1^{(2)}+4 m^4-9\right)\,z^6+O (z^7).
\end{equation}
Using this expansion, we can derive the following asymptotic form of the metric
\begin{equation}
\mathrm{d}s^2=\frac{1}{z^2}\big[\mathrm{d}s_{0}^2+\mathrm{d}s_2^2\,z^2+\mathrm{d}s_4^2\,z^4+ \ldots \big],
\end{equation}
with
\begin{subequations}
	\begin{align}
	&\mathrm{d}s_{0}^2=-\mathrm{d}t^2+\mathrm{d}w_1^2+\mathrm{d}w_2^2+\mathrm{d}w_3^2,
	\\
	&\mathrm{d}s_{2}^2=-\frac{m^2}{3} \Big(-\mathrm{d}t^2+\mathrm{d}w_1^2+\mathrm{d}w_2^2+\mathrm{d}w_3^2\Big),
	\\
	&\mathrm{d}s_{4}^2=\frac{1}{4} \Big(2 m q_2^{(1)}-3 q_1^{(2)}+3\Big)\mathrm{d}t^2+\frac{1}{4} \Big(1-2 m q_2^{(1)}-q_1^{(2)}\Big) \big(\mathrm{d}w_1^2+\mathrm{d}w_2^2+\mathrm{d}w_3^2\big).
	\end{align}
\end{subequations}
Similarly, for the scalar fields we find
\begin{subequations}
	\begin{align}
	&\mu = m\,z+\bigg(q_2^{(1)}+\frac{m^3}{9}\bigg)\,z^3 +O(z^4),
	\\
	&\sigma = q_3^{(0)}\,z^3 +O(z^4).
	\end{align}
\end{subequations}
Following \cite{Bianchi:2001kw,Bianchi:2001de} we can now compute the stress energy tensor and expectation values. Let $\langle\mathcal{O}_3\rangle$ and $\langle\mathcal{O}_4\rangle$ be the expectation values of the operators defined in \eqref{fermion bilinears}, dual to $\mu$ and $\sigma$, respectively. Then we have
\begin{subequations}
	\label{eqs:asymp}
	\begin{equation}
	\langle \mathcal{O}_3\rangle =-\frac{q_2^{(1)}}{2\pi G_5} \,, 
	\end{equation}
	\begin{equation}
\langle \mathcal{O}_4\rangle =-\frac{q_3^{(0)}}{2\pi G_5} \,,
	\end{equation}
\end{subequations}
and the holographic stress tensor is	
\begin{equation}
\langle T_{\mu\nu}\rangle \, \mathrm{d}x^\mu \mathrm{d}x^\nu = \frac{1}{16\pi G_5} \left[ \Big(3-3q_1^{(2)}-2\,m\,q_2^{(1)} \Big) \mathrm{d}t^2 +  \Big(1-q_1^{(2)}+2\,m\,q_2^{(1)}\Big)\sum_{i=1}^3 \mathrm{d} w_i^2 \right].
\end{equation}
As expected from Ward identities \cite{Bianchi:2001kw,Bianchi:2001de}, this holographic stress tensor is not traceless. Instead, we have
\begin{equation}
\langle T^{\mu}_{\phantom{\mu}\mu}\rangle +m\langle \mathcal{O}_3\rangle=0.
\end{equation}
Finally, from (\ref{eqs:asymp}) we can read off the holographic energy density:
\begin{equation}
\rho = \frac{1}{16\pi G_5} \Big(3-3q_1^{(2)}-2\,m\,q_2^{(1)}\Big),
\label{eq:holorho}
\end{equation}
and, as expected, $\rho = 0$ when evaluated on the \emph{supersymmetric} solutions \eqref{eq:super1} and \eqref{eq:super2}.

%%%%%%%%%%%%%%%%%%%%%%%%%%%%%%%%%%%%%
\subsubsection{Temperature and entropy}
%%%%%%%%%%%%%%%%%%%%%%%%%%%%%%%%%%%%%

We want to determine the entropy density and the temperature of the holographic states dual to our supergravity solutions. Using the metric \emph{ansatz} \eqref{eq:ansatz5D}, the entropy density is
\begin{equation}
s=\frac{A}{4\,\Delta w_1\,\Delta w_2\,\Delta w_3\,G_5},
\end{equation}
where $A$ is the area of the intersection of the horizon with a partial Cauchy surface of constant $t$ and we take the spatial boundary directions $w_i$ to be periodic with periods $\Delta w_i$.  Under the gauge fixing described below \eqref{eq:ansatz5D}, the horizon area is just $A = \Delta w_1\,\Delta w_2\,\Delta w_3$. This yields the entropy density
\begin{equation}
s=\frac{1}{4G_5}.
\end{equation}
To obtain the temperature $T$, we compute the surface gravity $\kappa = 2\pi T$ via
\begin{equation}
\nabla_a(k^bk_b) \Big\rvert_{\mathcal{H}}=-2 \kappa \,k_a,
\end{equation}
where the horizon at $y=1$ is a Killing horizon of $k^a=(\partial_t)^a$. We first change to ingoing Eddington-Finkelstein coordinates via
\begin{equation}
\mathrm{d}t=\mathrm{d}v+\frac{1}{2\sqrt{y\,f(y)g(y)}}\, \mathrm{d}y=\mathrm{d}v+\frac{1}{2(1-y^2)\sqrt{y\,q_1(y)q_4(y)}}\,\mathrm{d}y,
\end{equation}
which brings the metric into the form
\begin{equation}
\mathrm{d}s^2_{\mathrm{EF}}=\frac{1}{y}\left[-(1-y^2)\mathrm{d}v^2-\frac{1}{\sqrt{y}}\sqrt{\frac{q_1}{q_4}}\,\mathrm{d}v\,\mathrm{d}y+\mathrm{d}w_1^2+\mathrm{d}w_2^2+\mathrm{d}w_3^2\right],
\end{equation}
Under this coordinate transformation, $k^a = (\partial_v)^a$ and $k_a \big\rvert_{\mathcal{H}}= -\frac{\sqrt{q_1(1)}}{2\sqrt{q_4(1)}} \delta^y_a$, consistent with the fact that $y=1$ is a null hypersurface.  Evaluating $\nabla_a (k^b k_b) \big\rvert_{y = 1}$ we finally obtain the temperature:
\begin{equation}
\label{eq:temperature}
 T=\frac{\sqrt{q_1(1)q_4(1)}}{\pi}.
\end{equation}

%%%%%%%%%%%%%
\subsubsection{Conformal thermodynamic invariants \label{sec:conformal}}
%%%%%%%%%%%%%
The quantities defined in the preceding section are not quite the physical ones.  In the UV, the boundary theory becomes conformal, and thus the only physically-meaningful quantities are conformal invariants.  To build these, we simply soak up any dimensions with the appropriate power of the temperature.  Thus the conformally-invariant entropy density is
\begin{subequations}
	\begin{equation}
	\widehat{s}=\frac{s}{T^3},
	\end{equation}
	and the conformally-invariant energy density, and expectation values $\langle \mathcal{O}_3\rangle$ and $\langle \mathcal{O}_4\rangle$ are given by
	\begin{align}
	&\widehat{\rho}=\frac{\rho}{T^4},
	\\
	&\langle \widehat{\mathcal{O}}_3\rangle=\frac{\langle \mathcal{O}_3\rangle}{T^3},
	\\
	&\langle \widehat{\mathcal{O}}_4\rangle=\frac{\langle \mathcal{O}_4\rangle}{T^3}.
	\end{align}
	We will also be interested in plotting the free energy density $f = \rho-T s$, which has the same conformal dimension as $\rho$, and so we should focus on studying
	\begin{equation}
	\widehat{f}=\frac{f}{T^4}=\widehat{\rho}-\widehat{s}.
	\end{equation}
\end{subequations}
All of these quantities should be a written as a function of the dimensionless mass deformation
\begin{equation}\label{confInvMdef}
\widehat{m} \equiv \frac{m}{\sqrt{3} T} \,,
\end{equation}
where the factor of $\sqrt{3}$ is inserted here for later convenience. We also note that we will set $G_5 = \pi/(2N^2)$, so that all our extensive thermodynamic quantities will appear normalised relative to $N^2$. Further note that it follows from \eqref{confInvMdef}  that having a small conformally invariant mass-deformation parameter $\widehat{m}$ corresponds to high temperature.

%%%%%%%%%%%%%%%%%%%%%%%%%%%%%%%%%%%%%
\subsection{Thermal phases of the 2-scalar GPPZ subsector: results}
%%%%%%%%%%%%%%%%%%%%%%%%%%%%%%%%%%%%%
\subsubsection{Thermal phases with $\sigma = 0$: no gaugino condensation \label{sec:NOgaugino}}
\begin{figure}
	\centering
	\includegraphics[width=\linewidth]{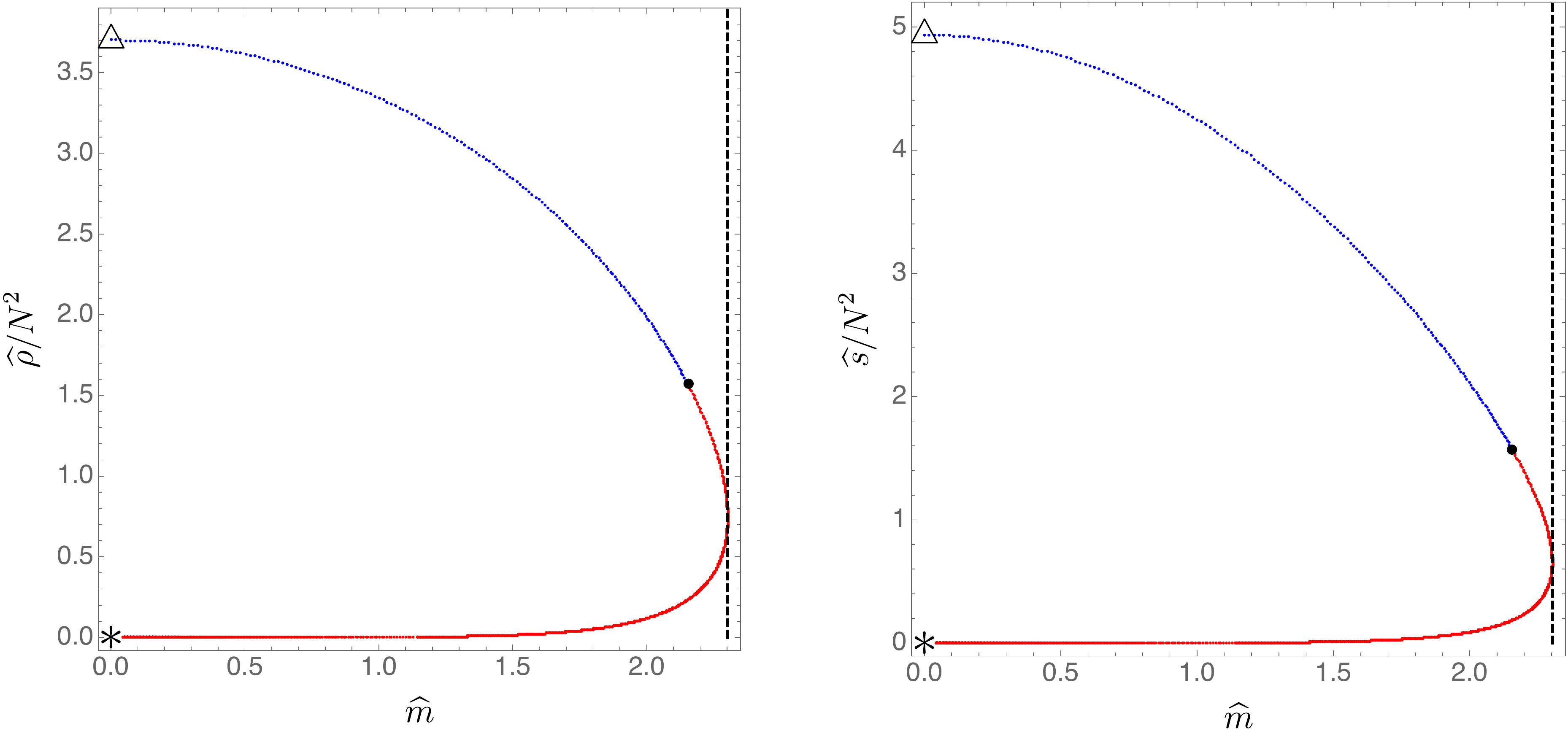}
	\caption{\label{fig:thermo1} Energy density $\widehat{\rho}$ ({\bf left panel}) and entropy density $\widehat{s}$ ({\bf right panel}) as a function of the mass deformation $\widehat{m}$ (solutions with no gaugino condensate, $\sigma=0$). The vertical dashed line intersects the turning point at $\widehat{m}=\widehat{m}_{t}\sim 2.30$; see \eqref{turningPoint}. The solution marked with a triangle $\triangle$ (with $\widehat{m}=0$, $\widehat{\rho}\neq 0$ and $\widehat{s}\neq 0$) corresponds to the  planar AdS$_5$ black hole. The asterisk $\ast$ (with $\widehat{m}=0$, $\widehat{\rho}=0$ and $\widehat{s}=0$) describes a singular solution whose Weyl tensor blows up (see Fig.~\ref{fig:thermo2}); we will give strong evidence that it describes the GPPZ solution (\ref{eq:super1}). The black disk $\bullet$ pinpoints the critical mass deformation $\widehat{m}=\widehat{m}_c\sim 2.15$, see \eqref{phaseTransition}, at which a (most liklely first-order) phase transition occurs (see Fig.~\ref{fig:canonical}). The blue dotted branch of solutions (from $\bullet$ to $\triangle$) dominates the canonical ensemble (since it has negative free energy density) while the red dotted branch  (from $\bullet$ into $\ast$)  has positive free energy density: see Fig.~\ref{fig:canonical}. These color and symbol codes will be kept through Figs. \ref{fig:thermo2}$-$\ref{fig:thermo5b}.}
\end{figure}

\begin{figure}[ht]
	\centering
	\includegraphics[height=0.45\linewidth]{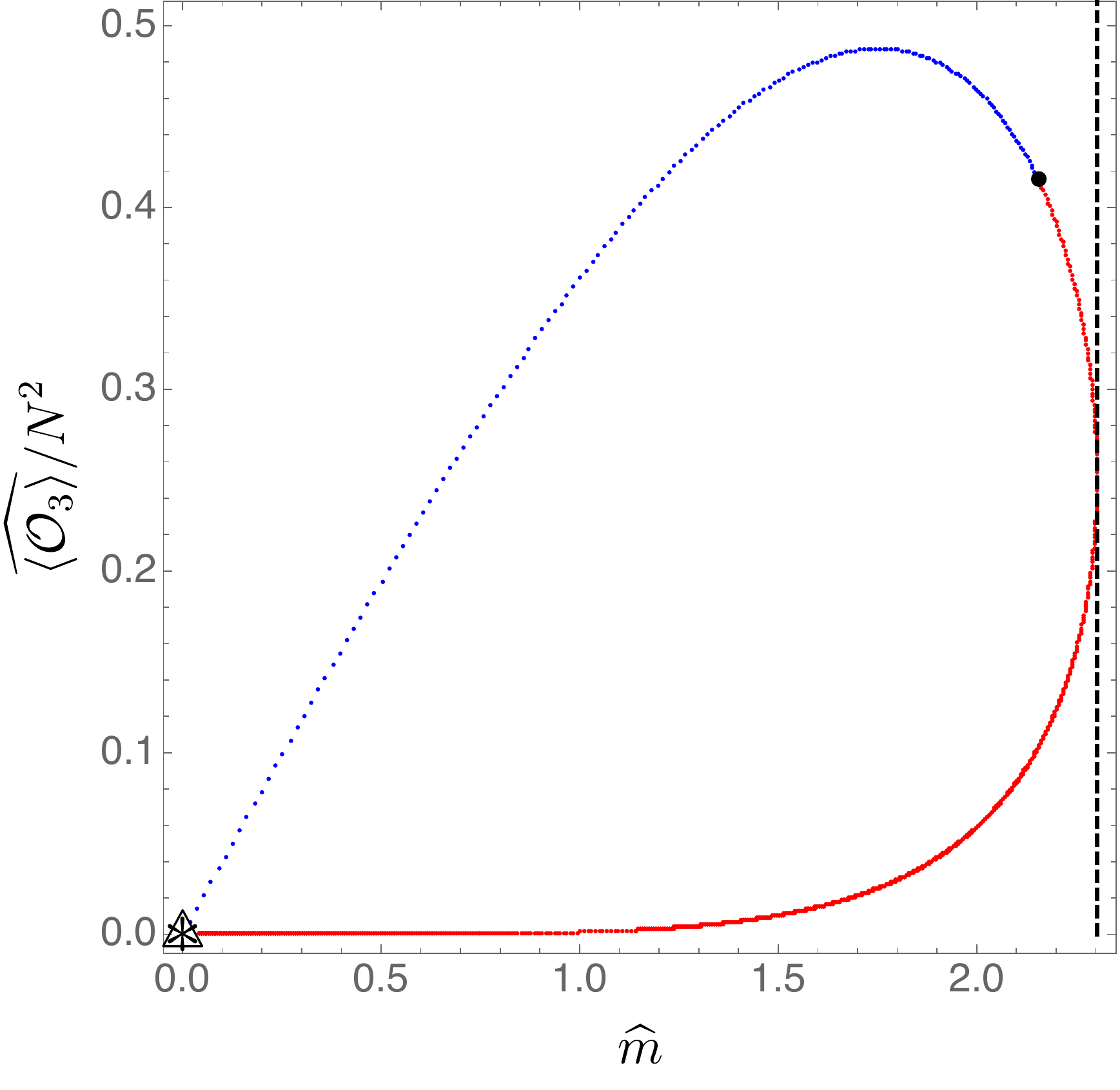}
	\hspace{0.5cm}
	\includegraphics[height=0.45\linewidth]{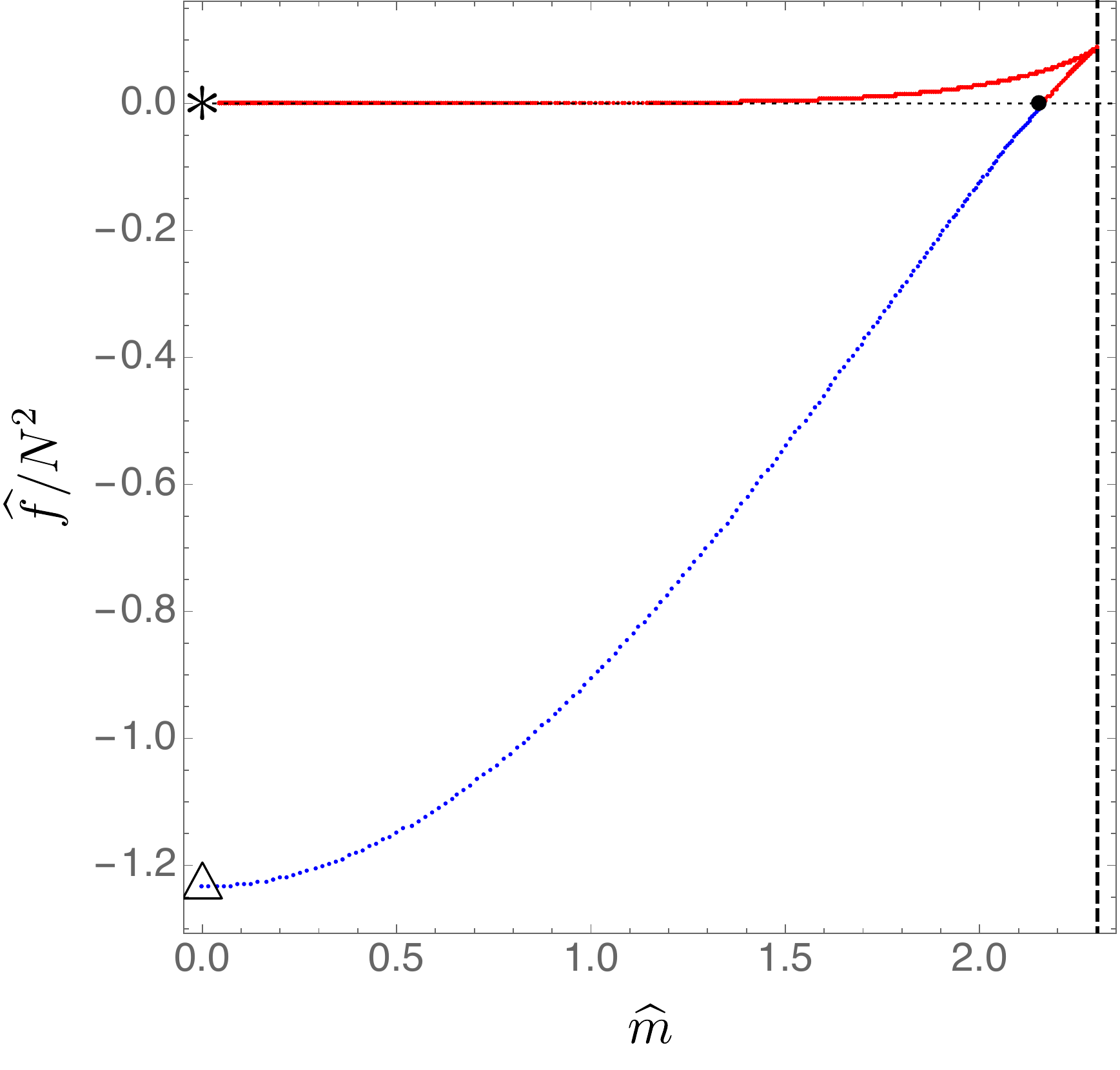}
	\caption{{\bf Left panel:} The vacuum expectation value $\langle \widehat{\mathcal{O}}_3\rangle$ plotted as a function of $\widehat{m}$. {\bf Right panel:} Canonical ensemble phase diagram: conformally-invariant free energy $\widehat{f}$ as a function of the conformally-invariant mass deformation $\widehat{m}$. At $\widehat{m}=\widehat{m}_c\sim 2.15$ (the solution denoted with a black disk $\bullet$ with $\widehat{f}=0$), a (most likely first-order) phase transition in the canonical ensemble occurs. Note that the $\ast$ solution (with $\widehat{\rho}=0$) has $\widehat{f}=0$, consistent with the claim that it describes the supersymmetric GPPZ solution (\ref{eq:super1}). Again, notice  we use the same ccolorolour and symbol codes detailed in the caption of Fig.~\ref{fig:thermo1}.}
	\label{fig:canonical}
\end{figure}

%%%%%%%%%%%%%%%%%%%%%%%%%%%%%%%%%%%%%
Solutions with nontrivial  gaugino $\sigma$ hair will only exist for sufficiently large values of the dimensionless mass deformation $\widehat{m}$, so we will first discuss the results for $\sigma=0$. In Fig.~\ref{fig:thermo1} we plot the energy density $\widehat{\rho}\,$ and entropy density $\widehat{s}\,$ of these solutions as a function of $\widehat{m}$. The first striking feature of both these figures is the presence of a turning point at (see the vertical dashed line) 
\begin{equation}\label{turningPoint}
\widehat{m}=\widehat{m}_{t}\approx 2.3029446\,.
\end{equation}
 For each value of $\widehat{m}<\widehat{m}_{t}$, we find two families of solutions with different values of $\widehat{\rho}$ and $\widehat{s}$. The family of solutions with lower energy and entropy densities are unstable against a Gregory-Laflamme instability along the brane directions $w_i$.  This follows from the analysis of \cite{Wald:2014bia,Schiffrin:2013zta}: the instability sets in precisely at the turning point $\widehat{m}=\widehat{m}_t$ of the canonical phase diagram (free energy density vs mass deformation; see the right panel of Fig.~\ref{fig:canonical}).

\begin{figure}
	\centering
	\includegraphics[height=0.45\linewidth]{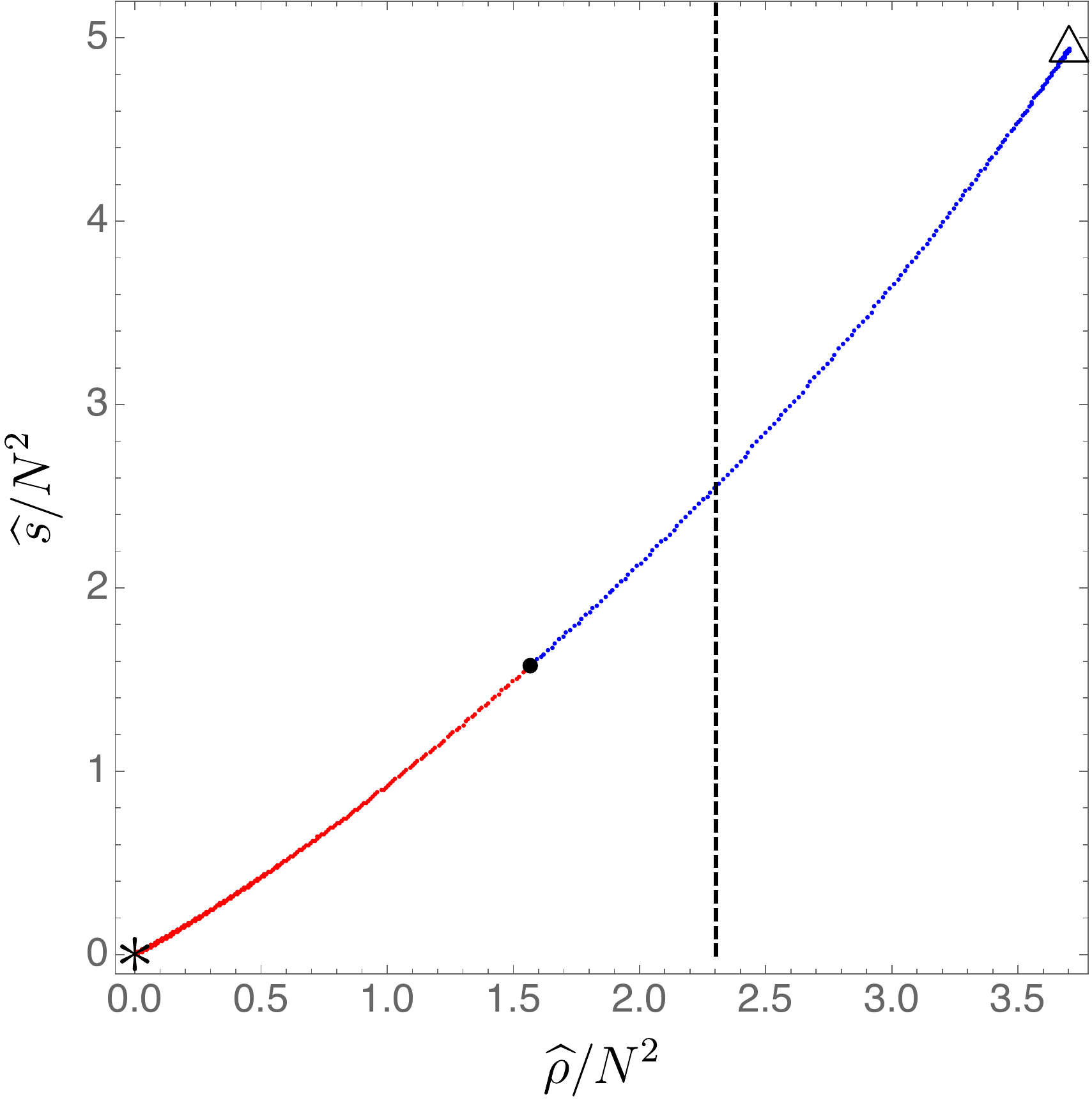}
	\hspace{0.5cm}
	\includegraphics[height=0.45\linewidth]{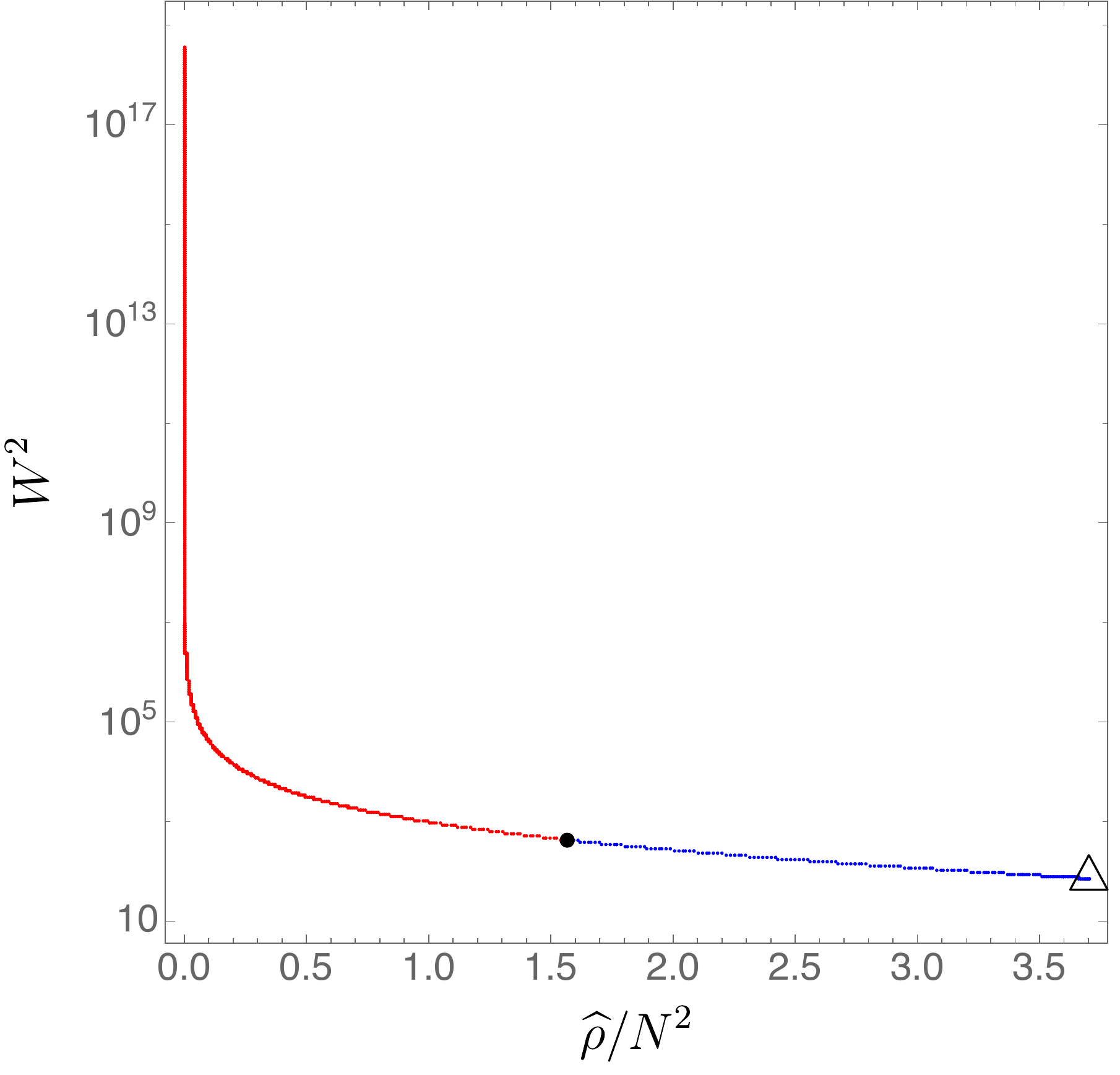}
	\caption{{\bf Left Panel:} Entropy density $\widehat{s}\,$ as a function of the energy density $\widehat{\rho}$. {\bf Right panel:} Logarithmic plot of the Weyl curvature $W^2\equiv C^{abcd}C_{abcd}|_{\mathcal{H}}$ as a function of $\widehat{\rho}$: a blow up seems to emerge as $\widehat{\rho}\to0$ (and $\widehat{m}\to 0$, $\widehat{s}\to 0$, {\it i.e.} as the $\ast$ solution of Fig.~\ref{fig:thermo1} is approached). Note that we use the same color and symbol codes detailed in the caption of Fig.~\ref{fig:thermo1}.}
	\label{fig:thermo2}
\end{figure}

\begin{figure}[ht]
	\centering
	\includegraphics[width=\linewidth]{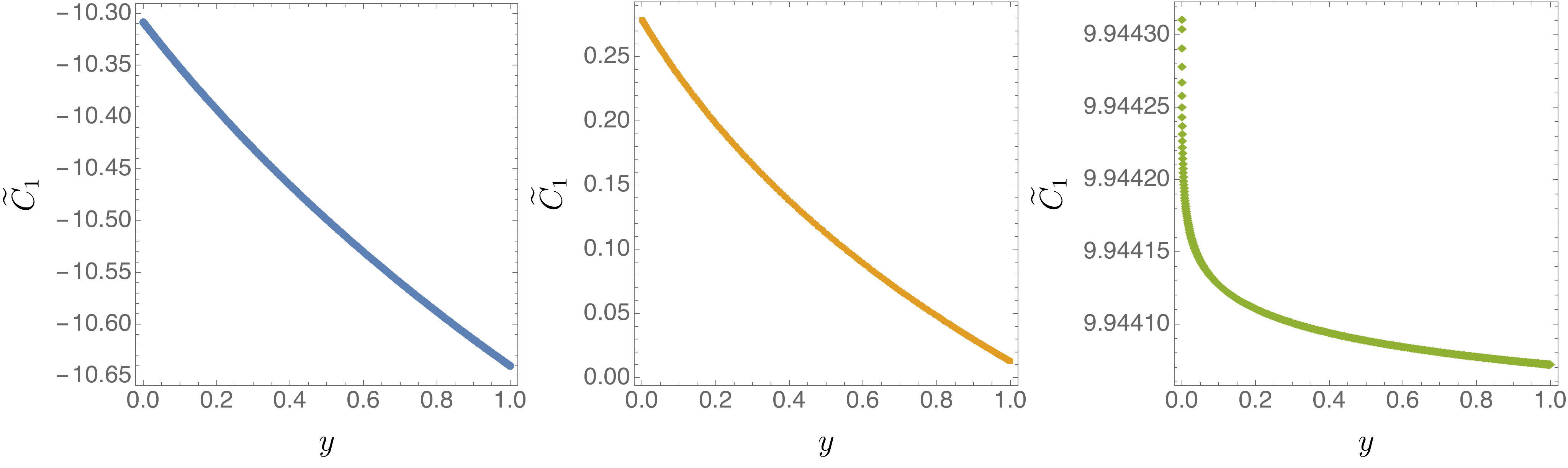}
	\caption{The plots of $\widetilde{C}_1(y)$ as a function of $y$, for different fixed values of $\widehat{\rho}$: from left to right, we have $\widehat{\rho}/N^2\approx 3.7, 1.2, 6\times 10^{-12}$. As $\widehat{\rho}\to 0$, $\widehat{m}$ is approaching a constant value. This gives evidence for the claim that the $\ast$ solution in our plots is the GPPZ supersymmetric solution (\ref{eq:super1}).}
	\label{fig:c1}
\end{figure}

\begin{figure}[ht]
	\centering
	\includegraphics[width=0.5\linewidth]{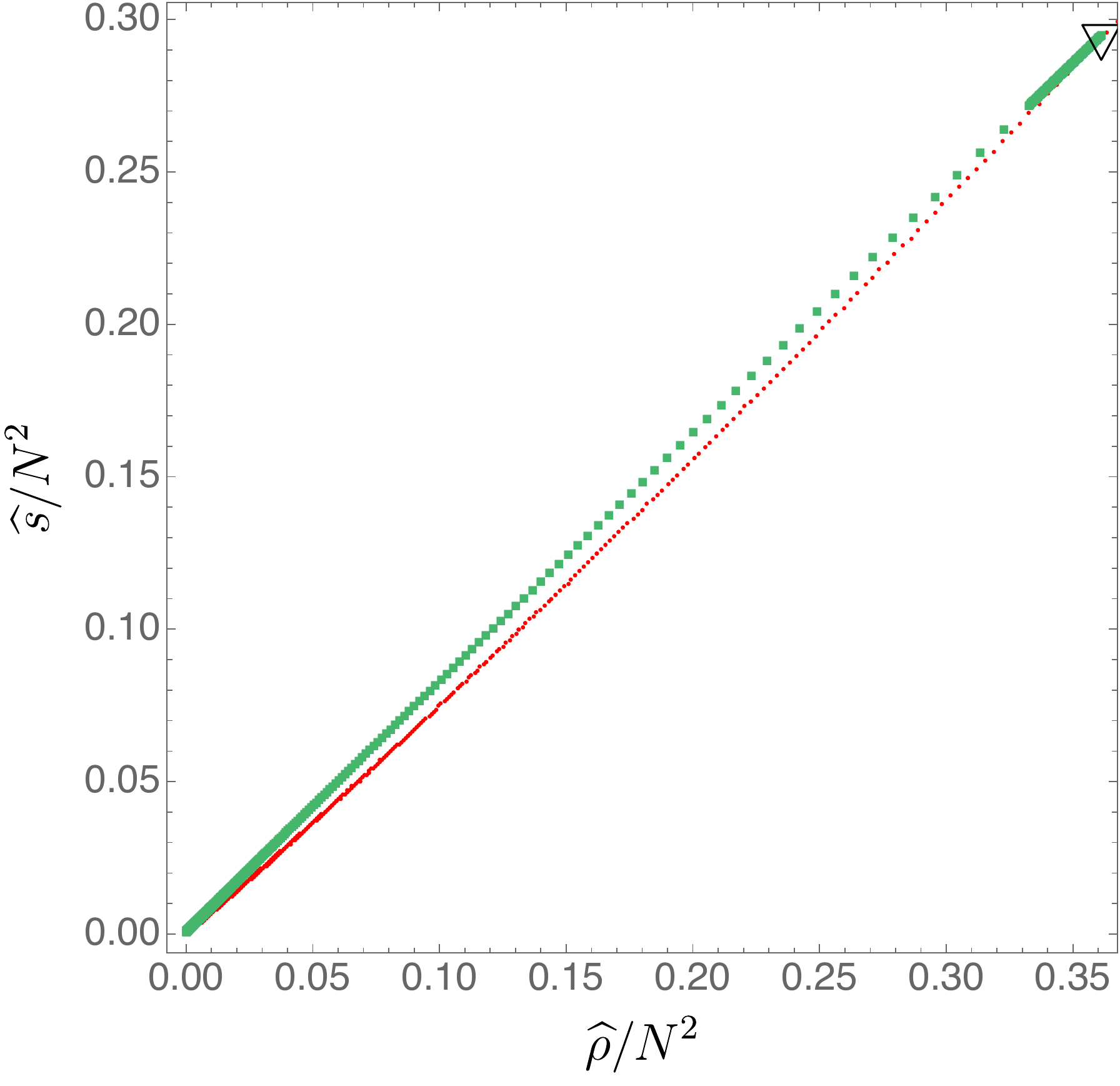}
	\caption{The entropy density $\widehat{s}/N^2$ as a function of $\widehat{\rho}/N^2$: the green  squares (upper branch) describes the family of solutions with a nontrivial gaugino condensate, $\sigma\neq 0$, while the red disks (lower branch) correspond to solutions with $\sigma=0$. Note that the latter family was constructed in Section \ref{sec:NOgaugino} and the associated results presented here represent a zoom-in of Fig.~\ref{fig:thermo2} in the relevant region $\widehat{m}\leq \widehat{m}_{\mathrm{g}}$ where it coexists with the gaugino condensate solutions.}
	\label{fig:thermo4}
\end{figure}

In the left planel of Fig.~\ref{fig:thermo2} we plot the entropy density $\widehat{s}\,$ as a function of energy density $\widehat{\rho}$, which is possible because the turning point $\widehat{m}=\widehat{m}_{t}$ is the same for the energy and entropy densities, so that $\widehat{s}\,$ is a monotonic function of $\widehat{\rho}$. In the right panel, we plot $W^2\equiv C^{abcd}C_{abcd}|_{\mathcal{H}}$, where $C$ is the Weyl tensor, as a function of $\widehat{\rho}$: we find that the solution becomes singular as $\widehat{\rho}\,$ decreases (this corresponds to the $\ast$ in Fig.~\ref{fig:thermo1}). So in the plots of Figs.~\ref{fig:thermo1}$-$\ref{fig:thermo5b} the asterisk $\ast$ (with $\widehat{m}=0$, $\widehat{\rho}=0$ and $\widehat{s}=0$) represents a singular solution.

Returning to Fig.~\ref{fig:thermo1}, there are two solutions with $\widehat{m} = 0$. The solution with finite energy and entropy density (identified by a triangle $\triangle$) corresponds to the (undeformed) planar AdS$_5$ black hole  \eqref{eq:ansatz5D}-\eqref{eq:5dplanarBH}. The other one, with ${\widehat{m} = \widehat{\rho} = \widehat{s} = 0}$ (pinpointed by an asterisk $\ast$ in our plots) is singular, as described in the previous paragraph. 

Naturally, one might wonder whether this singular solution $\ast$ corresponds to the GPPZ solution in (\ref{eq:super1}). We find several bits of numerical evidence that this is true. Indeed, we first note that the following combination of functions is constant for the GPPZ solution (\ref{eq:super1}):
$$
\log\left[\sinh ^2\left(\frac{\mu(\tilde{y})}{\sqrt{3}}\right) e^{2 \tilde{\phi} \left(\tilde{y}\right)}\right]=C_1\,.
$$
We next note that $e^{2\tilde{\phi}(\tilde{y})}$ can be made gauge invariant via the identification 
\begin{equation}
{g_{w_i w_i}=e^{2\tilde{\phi}(\tilde{y})}=1/y}, 
\end{equation}
where $w_i$ are Killing directions. Thus, to support our claim, in our numerical solution we must test whether the combination $\widetilde{C}_1(y) = 2\log\left[q_2(y)/\sqrt{3}\right]$ becomes a constant as we decrease $\widehat{\rho}$ (recall the relation \eqref{mu-q2} between $\mu$ and $q_2$). In Fig.~\ref{fig:c1} we  plot $\widetilde{C}_1(y)$ as a function of $\widehat{m}$ and find that it approachs a constant value as $\widehat{\rho}$ decreases towards  zero (equivalently, as $\widehat{s}\,$ approaches  zero since they are monotonically related). Further evidence that $\ast$ in our plots describes the GPPZ solution is given shortly.\footnote{It follows from \eqref{confInvMdef}  that a small conformally-invariant  mass deformation $\widehat{m}$ corresponds to a high temperature $T$ and finite $m$. One of our branches of thermal solutions approaches,  in the $\widehat{m}\to 0$ limit, the GPPZ solution  (the $\ast$ solution in our plots). Having a thermal solution approaching the confined supersymmetric vacuum of the theory in the $T\to\infty$ limit is a feature of many systems. As a simple example, recall that $AdS$ is a supersymmetric solution and the thermal solution of the theory with vanishing matter fields (the planar AdS-Schwarzschild solution) approaches the confined AdS vacuum as $T\to \infty$.
}

Another quantity of interest to interpret our solutions is $\langle \widehat{\mathcal{O}}_3\rangle$, the conformal invariant constructed from the expectation value of the chiral superfields, see \eqref{fermion bilinears}, 
\begin{equation}
\langle \widehat{\mathcal{O}}_3 \rangle = \frac{1}{T^3}\sum_{i=1}^3 \langle \text{tr} \left(  \lambda^i \lambda^i \right) \rangle \,,
\end{equation}
which we plot in the left panel of Fig.~\ref{fig:canonical} as a function of $\widehat{m}$. The fact that $\langle \widehat{\mathcal{O}}_3\rangle$ vanishes as $\widehat{\rho}\to0$ also agrees with the expectations about the supersymmetric theory, which are confirmed even in the singular GPPZ solution of (\ref{eq:super1}). (Notice that since we are looking for solutions with $\sigma = 0$, the gaugino condensate vanishes, $\langle \widehat{\mathcal{O}}_4 \rangle = 0$).

Of course, the central quantity of interest for the physical interpretation of our solutions is the free energy. In the right panel of Fig.~\ref{fig:canonical} we plot the conformally-invariant free energy density $\widehat{f}$ as a function of  the mass deformation density $\widehat{m}$. Recall that supersymmetric solutions calibrate the free energy (they have $\widehat{f} = 0$). Hence, we expect a phase transition between the high-temperature deconfined phase dual to our solution and the supersymmetric vacua of our theory at the point where $\widehat{f}$ crosses zero and becomes negative. This occurs at $\widehat{m}=\widehat{m}_c$ with
\begin{equation}\label{phaseTransition}
\widehat{m}_c\approx 2.15467491205(6)\,.
\end{equation}
Since five-dimensional supergravity does not seem powerful enough to capture the supersymmetric vacua of $\mathcal{N} = 1^*$ theory, it is conceivable that our analysis here is also missing some possible intermediate-temperature phases.  These phases might correspond to solutions captured by the 5d description by turning on other operators within the $SO(3)$-invariant truncation, such as $\cO_1$, or $\cO_2$.  There could also be phases that break $SO(3)$ invariance (although this should not be important for questions regarding the expected smooth Polchinski-Strassler vacua).  Or there may be phases that are captured only by the 10d supergravity description. In the absence of precise information about these phases we cannot confidently ascertain the order of this transition.

If this phase transition leads directly from the high-temperature phase to the supersymmetric solution (without any intermediate phases), then it is first order, since the condensate $\langle \widehat{\mathcal{O}}_3 \rangle$ discontinuously jumps from zero to a non-zero value as $\widehat{m}$ is lowered. It may also be identified as a `deconfinement' transition\footnote{Even if strictly-speaking the low-temperature phases can be either confining, screening of oblique confining.} since the (un-normalized) free energy $F$ jumps from 0 to $\mathcal{O}(N^2)$.\footnote{Implicit in the normalizations of Sec.~\ref{sec:conformal} is that $F \propto N^2 \widehat{f}$, so that any $\mathcal{O}(1)$ value of $\widehat{f}$ corresponds to a free energy which scales as $N^2$. } 
We will come back to this question in the end of Section \ref{sec:IIBresults} after we construct in this section the full ten-dimensional solution dual to the thermal phase of mass-deformed theory, and discuss the possibility of having `ringoid' thermal phases.

%%%%%%%%%%%%%%%%%%%%%%%%%%%%%%%%%%%%%
\subsubsection{Thermal phases with $\sigma \neq 0$: gaugino condensation}
%%%%%%%%%%%%%%%%%%%%%%%%%%%%%%%%%%%%%

\begin{figure}[ht]
	\centering
	\includegraphics[width=\linewidth]{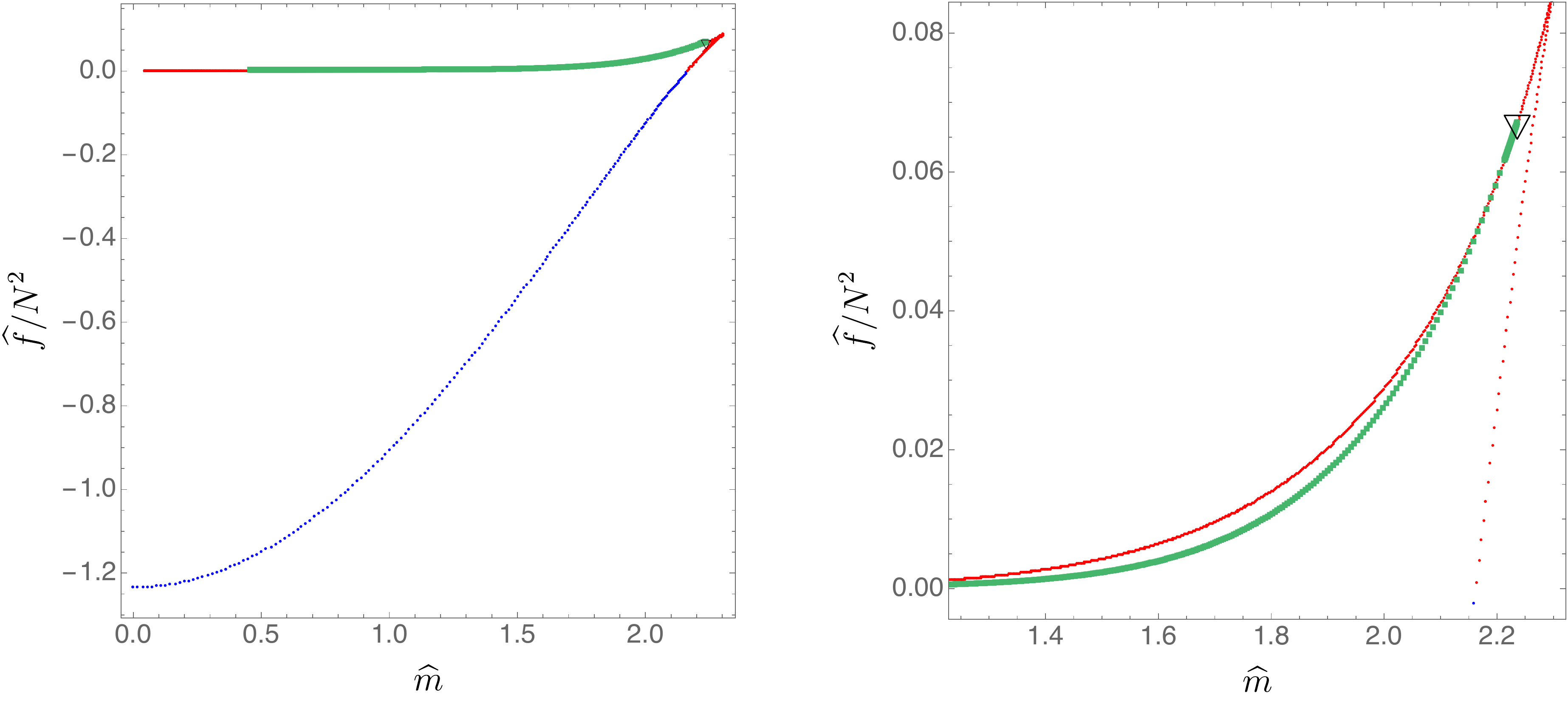}
	\caption{Canonical-ensemble phase diagram $\widehat{f}/N^2$ vs $\widehat{m}$, this time also with the gaugino condensate ($\sigma\neq0$) family of solutions. The blue and red disks describe the solutions with $\sigma=0$ already presented in Fig.~\ref{fig:canonical}. But this time we also display, using  green squares, the branch of solutions with $\sigma\neq0$. The black inverted triangle $\triangledown$ corresponds to $\widehat{m}=\widehat{m}_{\mathrm{g}}\sim 2.24$; see \eqref{gauginoMass}. The right panel zooms close to this second-order transition point between the two phases. The nonzero-gaugino branch of solutions never dominates the canonical ensemble.}
	\label{fig:thermo5a}
\end{figure}

In this section we want to take a step further and construct solutions with a gaugino condensate, $\sigma \neq 0$. As shown in Fig.~\ref{fig:thermo4}, we find that this family branches off from the solutions with $\sigma=0$ (of the previous section) at a particular value $\widehat{m}=\widehat{m}_{\mathrm{g}}$  with (see the  black inverted triangle $\triangledown$)
 \begin{equation}\label{gauginoMass}
\widehat{m}_{\mathrm{g}}\approx 2.2355679948.
\end{equation} 
Below this critical mass deformation, the $\sigma\neq 0$ and $\sigma=0$ solutions coexist. In this region, for a given energy density, the thermal solutions with $\sigma \neq 0$ have higher entropy density than the solutions with $\sigma=0$. The transition at $\widehat{m}=\widehat{m}_{\mathrm{g}}$ between these two phases is thus second order. 

\begin{figure}[ht]
	\centering
	\includegraphics[width=\linewidth]{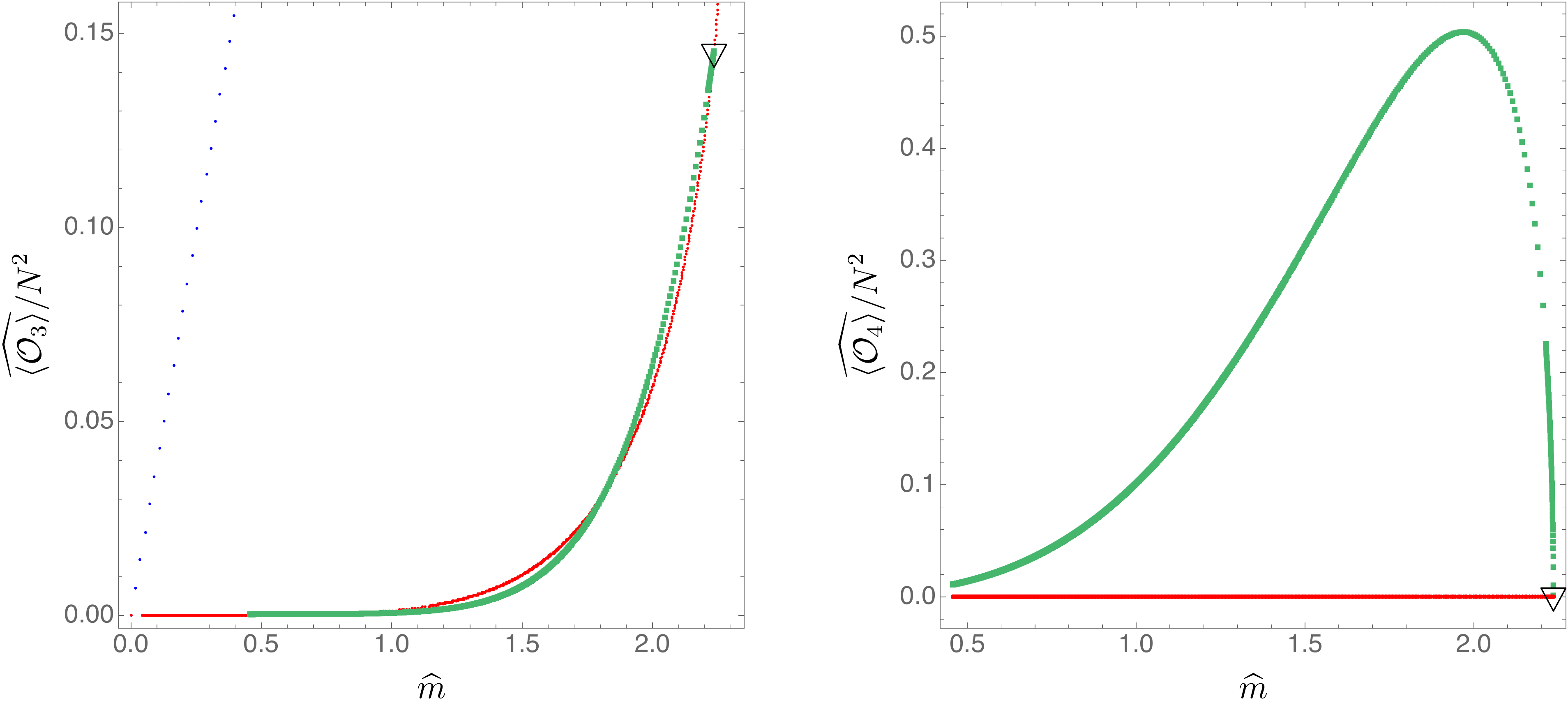}
	\caption{The expectation values $\langle \widehat{\mathcal{O}}_3\rangle$ (left panel) and $\langle \widehat{\mathcal{O}}_4\rangle$ (right panel) as a function of $\widehat{m}$. The red and blue dotted solutions have $\sigma=0$ (and thus $\langle \widehat{\mathcal{O}}_4\rangle=0$), and were already displayed in the left panel of Fig.~\ref{fig:canonical}: here we zoom in the region of interest. On the other hand, the green squares represent the gaugino-condensate family of solutions with $\sigma\neq 0$ hence $\langle \widehat{\mathcal{O}}_4\rangle\neq 0$. The black inverted triangle $\triangledown$ denotes the point of the merger of the two families at $\widehat{m}=\widehat{m}_{\mathrm{g}}$.}
	\label{fig:thermo5b}
\end{figure}

\begin{figure}[hb]
	\centering
	\includegraphics[width=0.5\linewidth]{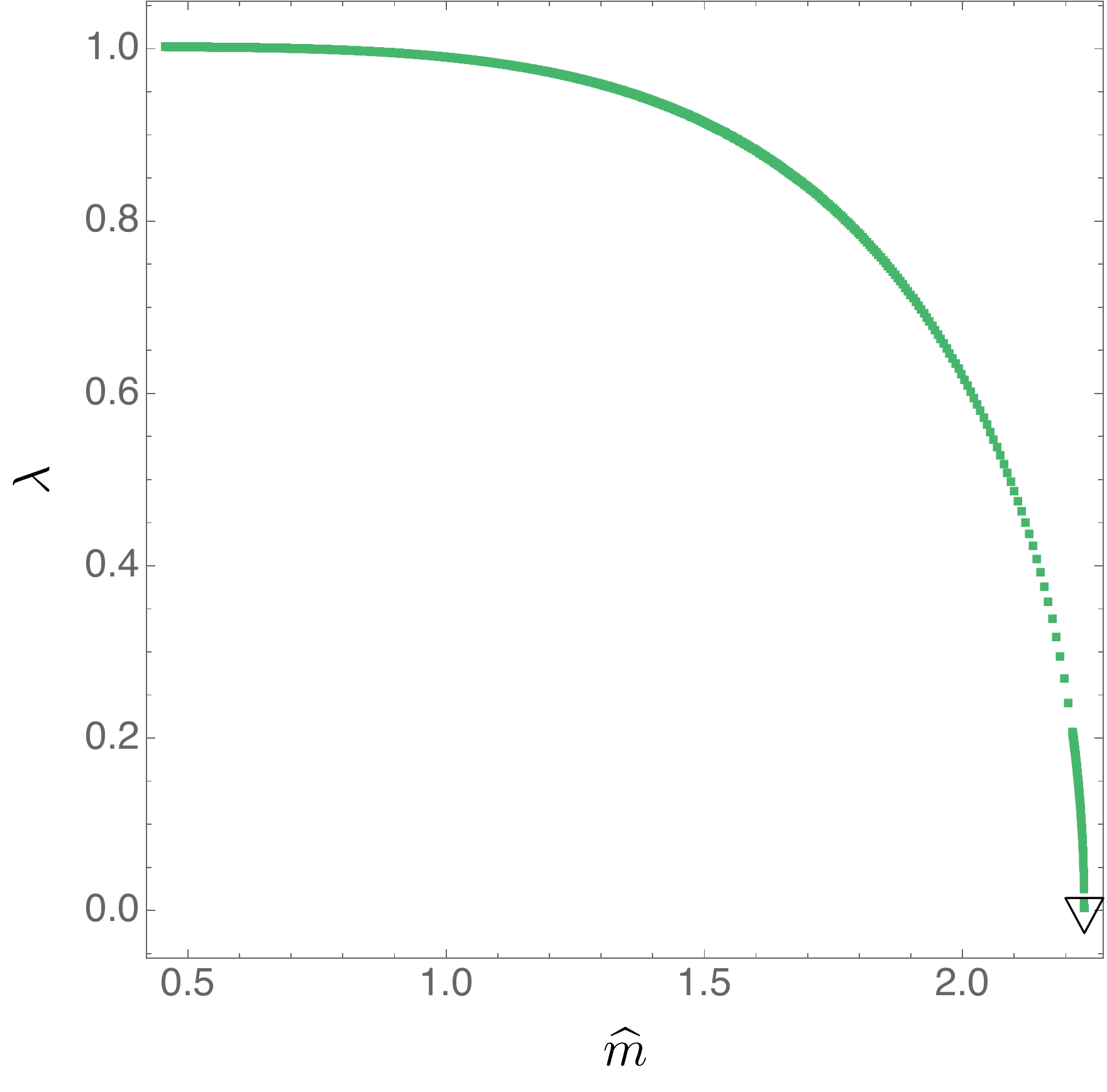}
	\caption{The dimensionless quantity $\lambda$, defined in \eqref{eq:lambda}, as a function of $\widehat{m}$: our solutions approach $\lambda\to1$ from below as $\widehat{m}\to0$ from above. Refs. \cite{Girardello:1999bd,Pilch:2000ue,Petrini:2018pjk,Bobev:2018eer} argue that physical flows must satisfy $\lambda \leq 1$.}
	\label{fig:thermo5c}
\end{figure}

As shown in Fig.~\ref{fig:thermo5a}, the free energy density of the gaugino-condensate solutions ($\sigma\neq 0$) is lower than that  of solutions with $\sigma=0$. But it is nevertheless positive. Hence, the gaugino-condensate phase never dominates the canonical ensemble.

For completeness, in Fig.~\ref{fig:thermo5b} we present $\langle \widehat{\mathcal{O}}_3\rangle$ and $\langle \widehat{\mathcal{O}}_4\rangle$ measured in the gaugino condensate phase and compare them with the expectations values of the phase with $\sigma=0$.

Finally, we analyze the behavior of $\lambda$, defined in~(\ref{eq:lambda}), as we dial $\widehat{m}$. In \cite{Girardello:1999bd} the authors argue that physical flows must satisfy $\lambda\leq1$. Furthermore, in \cite{Pilch:2000fu,Petrini:2018pjk,Bobev:2018eer} it was argued that the singularity of the ten-dimensional solution strongly depends on $\lambda$. In particular, for $\lambda<1$ the singularity appears ring-like, whereas for $\lambda=1$ it appears more like a point singularity. We find that our finite-temperature solutions do satisfy the aforementioned bound and in fact that $\lambda\to1$ from below as $\widehat{m}\to0$ from above (see Fig.~\ref{fig:thermo5c}). At present we have no understanding of why this happens.

%%%%%%%%%%%%%%
\section{The $SO(3) \times U(1)$-invariant finite-temperature flows in type IIB supergravity}
\label{sec:IIBflows}
%%%%%%%%%%%%%%

In this section we will construct solutions dual to thermal states in the $\mathcal{N}=1^*$ theory from the 10-dimensional perspective.  We will also present evidence that for high temperatures some of the ten-dimensional solutions we construct are uplifts of some of the five-dimensional gauged supergravity flows we constructed in Section~\ref{sec:5d}. The ten-dimensional solutions that would result from uplifting the 2-scalar GPPZ truncation should in general be cohomogeneity 3, which is beyond the scope of our numerical investigation. However, in the limit $\sigma = 0$ the isometry is enhanced to $SO(3) \times U(1)$, and the problem of finding the ten-dimensional solution problem can be formulated as cohomogeneity-2. We will construct this solution numerically, analyze its physics, and show that it agrees with the physics of the five-dimensional $\sigma = 0$ solution of Section~\ref{sec:5d}. This agreement gives us hope that five-dimensional supergravity may be more powerful than previously thought for capturing high-temperature physics correctly even for less-symmetric theories.

%%%%%%%%%%%%%
\subsection{The type IIB equations of motion}
%%%%%%%%%%%%%

Rather than working in `string theory conventions', we prefer to work with `supergravity conventions', which are a formulation of type IIB supergravity where the $SO(2)$ part of the $SL(2,\RR)$ symmetry is manifest \cite{Schwarz:1983qr,Pilch:2000ue,Grana:2001xn}.  The fundamental fields are a complex scalar $B$ from which we define the 1-forms
\begin{equation} \label{PQ def}
P \equiv f^2 \dd B, \qquad Q \equiv f^2 \Im(B \dd \bar B), \qquad f \equiv \frac{1}{\sqrt{1-|B|^2}},
\end{equation}
as well as a complex 2-form potential $A_2$, and a real 4-form potential $C_4$, whose field strengths are given by
\begin{align}
G_3 &\equiv f \big( F_3 - B \bar F_3 \big), \qquad \qquad \qquad F_3 \equiv \dd A_2, \\
F_5 &\equiv \dd C_4 + \frac{i}{16} \big( A_2 \wedge \bar F_3 - \bar A_2 \wedge F_3 \big).
\end{align}
These relations in turn imply the Bianchi identities
\begin{subequations}
	\label{eqs:bianchi}
	\begin{align}
	& \dd P - 2i \, Q \wedge P = 0, & & \dd Q + i \, P \wedge \bar P = 0, \\
	& \dd G_3 - i \, Q \wedge G_3 + P \wedge \bar G_3 = 0, & & \dd F_5 - \frac{i}{8} \, G_3 \wedge \bar G_3 = 0.
	\end{align}
\end{subequations}
The action in terms of these fields\footnote{The one-form $Q$ does not appear in the action, but appears in the equations of motion because of the complicated way in which $B$ is packaged into the various other fields.} is
\begin{equation} \label{IIB action}
\cS = \frac{1}{2 \kappa_{10}^2} \int \bigg( \hodge \cR - 2 \, P \wedge \hodge \bar P - \frac12 G_3 \wedge \hodge \bar G_3 - 4 \, F_5 \wedge \hodge F_5 - i \, C_4 \wedge G_3 \wedge \bar G_3 \bigg),
\end{equation}
where one must also impose the self-duality condition $F^5 = \hodge F^5$.  We note that the $U(1)$ part of the $SL(2, \RR)$ symmetry of type IIB acts on these fields via the global phase rotations
\begin{equation} \label{phase rotation}
A_2 \to e^{i \alpha} A_2, \qquad G_3 \to e^{i \alpha} G_3, \qquad B \to e^{2i \alpha} B, \qquad P \to e^{2i \alpha} P,
\end{equation}
leaving all other fields invariant.  This simple expression of the $U(1)$ will be of great convenience in finding solutions.

The relation between  `supergravity fields/conventions' we use here and the often-used `string theory  fields/conventions' is detailed in  \cite{Grana:2001xn} and is reviewed next. One first combines the axion and dilaton into the complex field 
\begin{equation}
\tau \equiv C_0 + i e^{-\Phi},
\end{equation}
which is related to $B$ via a fractional linear transformation
\begin{equation}
B = \frac{1 + i\tau}{1 - i\tau}.
\end{equation}
This maps the upper half plane onto the unit disk, and thus the $SL(2,\RR)$ of the type IIB `string theory  (ST)  fields/conventions' onto $SU(1,1)$ (to which it is isomorphic).  One can also package the RR and NS-NS 3-forms of the `ST fields/conventions'  into a single complex 3-form,
\begin{equation}
G_3^{\text{ST}} \equiv F_3^{\text{ST}} - \tau H_3^\text{ST}.
\end{equation}
Then the `ST fields/conventions' $\{G_3^\text{ST}, F_5^\text{ST}\}$ are related to the `supergravity fields/conventions' for $\{G_3, F_5\}$  that we use via \cite{Grana:2001xn}\footnote{Further note that Polchinski-Strassler (PS) \cite{Polchinski:2000uf} define $G_3^\text{PS}=\bar{G}_3$. Additionally, recall that Newton's constant (supergravity coupling) is related to the string theory fundamental scales via $2 \kappa_{10}^2=(2\pi)^7g_s^2\ell_s^8$, where $\ell_s=\sqrt{\alpha'}$ is the string length.} 
\begin{equation}
G_3 \equiv i \frac{g_s}{\kappa_{10}} \bigg( \frac{1+i \bar \tau}{1-i\tau} \bigg)^{1/2} \big( \Im \, \tau \big)^{-1/2} \, G_3^\text{ST}, \quad\qquad F_5 \equiv \frac{g}{4 \kappa_{10}} F_5^\text{ST},
\end{equation}
where $g_s$ is the string coupling (the value of $e^\Phi$ at infinity).  Finally, our metric $g_{ab}$ is just the Einstein-frame metric.  

The equations of motion which derive from the action \eqref{IIB action} are
\begin{subequations}
	\label{eqs:IIB}
	\begin{align}
	&R_{ab} = T^{(1)}_{ab}+T^{(3)}_{ab}+T^{(5)}_{ab},
	\label{eq:einstein}
	\\
	& \dd \hodge P - 2i \, Q \wedge \hodge P + \frac14 \, G_3 \wedge \hodge G_3 = 0,
	\label{eq:p1}
	\\
	& \dd \hodge G_3 - i \, Q \wedge \hodge G_3 - P \wedge \hodge \bar G_3 + 4i \, G_3 \wedge F_5 = 0,
	\label{eq:g3}
	\\
	& \dd \hodge F_5 - \frac{i}{8} \, G_3 \wedge \bar G_3 = 0,
	\label{eq:f5}
	\end{align}
\end{subequations}
where the terms on the right-hand side of \eqref{eq:einstein} are given by
\begin{subequations}
	\label{eq:einstein sources}
	\begin{align}
	&T^{(1)}_{ab}=P_a\bar{P}_b+P_b\bar{P}_a,
	\\
	&T^{(3)}_{ab}=\frac{1}{8} \Big[ (G_3)_a{}^{cd}(\bar{G}_3)_{bcd}+(G_3)_b{}^{cd}(\bar{G}_3)_{acd}-\frac{1}{6}g_{ab} (G_3)^{cde}(\bar{G}_3)_{cde} \Big],
	\\
	&T^{(5)}_{ab}=\frac{1}{6}(F_5)_{a}{}^{cdef}(F_5)_{bcdef}.
	\end{align}
\end{subequations}
Note in particular that all of the equations are invariant under the \emph{constant} phase rotation \eqref{phase rotation}.

%%%%%%%%%%%%%
\subsection{The cohomogeneity-2 \emph{ansatz}}
\label{sec:cohomo 2}
%%%%%%%%%%%%%

As pointed out earlier, the uplift metric \eqref{uplift metric} for the supersymmetric GPPZ flows acquires an extra $U(1)$ isometry in the $\sigma = 0$ limit.  Thus one might hope to study finite-temperature flows with $\sigma = 0$ using a cohomogeneity-2 \emph{ansatz}.  We will show that this is indeed possible.

First, let us discuss the boundary conditions.  According to \cite{Polchinski:2000uf}, the fermion mass deformation \eqref{W deformation} should correspond to a particular non-normalizable mode in the complex 3-form $G_3$. If one chooses complex coordinates in the asymptotic $\RR^6 \simeq \CC^3$ transverse to the D3-branes:
\begin{equation} \label{complex coords}
z^1 \equiv x^1 + i x^2, \qquad z^2 \equiv x^3 + i x^4, \qquad z^3 \equiv x^5 + i x^6,
\end{equation}
then the 3-form should go asymptotically as
\begin{equation}
\label{eq:G3PS}
G_3 \sim \frac{\text{const}}{r^4} \Big( T_3 - \frac{4}{3} V_3 \Big) = \text{const} \times \dd \Big( r^{-4} S_2 \Big),
\end{equation}
where $T_3$ is a constant antisymmetric tensor containing the masses, 
\begin{equation}
T_3 \equiv m_1 \dd \bar z^1 \wedge \dd z^2 \wedge \dd z^3 + m_2 \dd z^1 \wedge \dd \bar z^2 \wedge \dd z^3 + m_3 \dd z^1 \wedge \dd z^2 \wedge \dd \bar z^3,
\end{equation}
and $V_3$ and $S_2$ are related to $T_3$ via
\begin{equation}
S_2 = \frac12 T_{mnp} x^m \dd x^n \wedge \dd x^p, \qquad T_3 = \frac13 \dd S_2, \qquad V_3 = \dd (\log r) \wedge S_2,
\end{equation}
where $r^2 = x^m x^m$. 

The backreaction of this 3-form is responsible for introducing dilaton, metric and five-form profiles that correspond to given equal masses to all the six bosons, such that the sum of the squares of the boson masses is equal to the sum of the squares of the fermion masses \cite{Bena:2015fev}. Since we consider a deformation with three equal fermion masses $m = m_1 = m_2 = m_3$, this will ensure that the three complex bosons will have the same mass as the fermions, and hence the bulk theory will be supersymmetric.\footnote{When the fermion masses are not equal one has to introduce extra bulk non-normalizable modes corresponding to traceless dimension-two boson bilinears to obtain a supersymmetric bulk theory \cite{Bena:2015fev}.}

Under the simultaneous $U(1)$ phase rotation  $z^i \to e^{i \alpha} z^i$ we  have
\begin{equation}\label{phase rotation2}
T_3 \to e^{i \alpha} T_3, \qquad S_2 \to e^{i \alpha} S_2, \qquad V_3 \to e^{i \alpha} V_3, \quad \text{and thus} \quad G_3 \to e^{i \alpha} G_3.
\end{equation}
Note that this phase rotation is exactly the $U(1)$ that appears in \eqref{u1 operators} acting on the operators $\cO_1, \cO_2$, and $\cO_4$.  Thus we see  that the 3-form should be \emph{charged} under this $U(1)$ even when those operators are turned off.

Note that the matter fields, as seen from  \eqref{phase rotation} and \eqref{phase rotation2}, are not invariant under the $U(1)$ phase rotations. However, under the $U(1)$ global phase rotations \eqref{phase rotation} the energy-momentum tensor $T_{ab}$ is invariant. Consequently, for solutions with $\sigma=0$, our metric and 4-form $C_4$ have enhanced  symmetry $SO(3) \times U(1)$ which is very welcome.

%%%%%%%%%%%%%
\subsubsection{Coordinates adapted to $SO(3) \times U(1)$ isometry}
%%%%%%%%%%%%%

We now want to write down a generic metric \emph{ansatz} with the desired $SO(3) \times U(1)$ isometry.  As a warm-up example, we first consider the metric of flat $\RR^6 \simeq \CC^3$ with the complex coordinates $z^1, z^2, z^3$ as defined in \eqref{complex coords}.  The $SO(3)$ rotates the $z^i$ among themselves as the \textbf{3} (fundamental) representation, whereas the $U(1)$ is the global phase $z^i \to e^{i \alpha} z^i$.  To obtain coordinates adapted to this symmetry, first observe that we can always write
\begin{equation}
z^i \equiv r (u^i + i v^i), \qquad u^i, v^i \in \RR^3, \qquad u^2 + v^2 = 1,
\end{equation} 
where $u^i$ and $v^i$ are two vectors in $\RR^3$ and correspond precisely to the $u^i, v^i$ in \eqref{uplift metric}.  Under the phase rotation $z^i \to e^{i \alpha} z^i$, one has
\begin{equation}
(u^i + i v^i) \to (\cos \alpha \, u^i - \sin \alpha \, v^i) + i (\sin \alpha \, u^i + \cos \alpha \, v^i),
\end{equation}
and thus
\begin{equation}
\vec u \cdot \vec v \to \cos (2\alpha) (\vec u \cdot \vec v) + \frac12 \sin(2\alpha) (u^2 - v^2).
\end{equation}
Under the constraint $u^2 + v^2 = 1$, each of the above terms is bounded between $\pm \frac12$, and therefore there exists a choice of phase $\alpha$ for which $\vec u \cdot \vec v \to 0$.  Thus if we pull out the phase $\alpha$ as a coordinate, we may take $u^i$ and $v^i$ to be orthogonal.  Next, by pulling out an overall $SO(3)$ rotation (with Euler angles $(\theta, \phi, \psi)$) of the $z^i$, we can always bring $u^i, v^i$ to lie in the 12 plane of $\RR^3$.  What remains is to satisfy the constraint $u^2 + v^2 = 1$ for two orthogonal vectors lying in the 12 plane, which introduces one last angle $\chi$.  We then obtain the expression
\begin{equation}
z^i = r e^{i \alpha} \big( \cos \chi \, R^i{}_1 + i \sin \chi \, R^i{}_2 \big),
\end{equation}
where $\alpha$ and $\chi$ are the coordinates we just discussed, and $R^i{}_1$ and $R^i{}_2$ are the first two columns of a rotation matrix $R^i{}_j \in SO(3)$ parameterized by the three Euler angles $(\theta, \phi, \psi)$ in the usual way:
\begin{equation}
R(\phi,\theta,\psi) = R_z(\phi) R_y(\theta) R_z(\psi).
\end{equation}
To write the metric, it is useful to define the usual Maurer-Cartan 1-forms $\sigma_i$ on $SO(3)$ via
\begin{equation}
	\sigma_i \equiv \frac12 \varepsilon_{ijk} (R^\top \dd R)_{jk},
\end{equation}
which under these conventions are \emph{right}-invariant, i.e. they satisfy
\begin{equation}
\dd \sigma_1 = \sigma_2 \wedge \sigma_3, \qquad \dd \sigma_2 = \sigma_3 \wedge \sigma_1, \quad \text{and} \quad \dd \sigma_3 = \sigma_1 \wedge \sigma_2.
\end{equation}
In these coordinates, the flat metric of $\RR^6$ becomes
\begin{equation} \label{R6 metric}
	\dd s_6^2 = \dd r^2 + r^2 \Big[ (\dd \alpha + \sin (2 \chi) \, \sigma_3)^2 + \dd \chi^2 + \cos^2 (2\chi) \, \sigma_3^2 + \sin^2 (\chi) \, \sigma_1^2 + \cos^2 (\chi) \, \sigma_2^2 \Big],
\end{equation}
where the five-sphere part of the metric is written in the form of a $U(1)$ bundle over $\CP^2$, although the $\CP^2$ is in unusual coordinates (we note that the $SO(3)$ isometry which we have made manifest is precisely the real part of the $SU(3)$ isometry of $\CP^2$).  However, for doing numerics we will find it useful to regroup the terms of the metric as:
\begin{equation}
\dd \Omega_5^2 = \dd \chi^2 + \sin^2 (\chi) \, \sigma_1^2 + \cos^2 (\chi) \, \sigma_2^2 + \cos^2(2\chi) \dd \alpha^2 + \big( \sigma_3^2 + \sin(2\chi) \dd \alpha \big)^2.
\end{equation}
In order to count each point of the 5-sphere exactly once, the coordinates should have the following ranges:
\begin{equation}
	\alpha \in [0, \pi), \qquad \chi \in [0, \tfrac{\pi}{4} ), \qquad \theta \in [0,\pi), \qquad \phi \sim \phi + 2\pi, \qquad \psi \sim \psi + 2\pi,
\end{equation}
which gives the correct area of the unit five-sphere, $\pi^3$.  The reason $\alpha$ is bounded in $[0,\pi)$ rather than identified modulo $2\pi$ is because a phase shift by $\alpha \to \alpha + \pi$ is equivalent to an $SO(3)$ rotation by $\pi$ in the 12 plane.  Similarly, the range of parameters for $\pi/4 < \chi < \pi/2$ is equivalent to the range $0 < \chi < \pi/4$ by an $SO(3)$ rotation which exchanges the axes 1 and 2.  Finally, for numerics it is more convenient to work with coordinates that appear algebraically in the metric, and thus we introduce
\begin{equation}
x \equiv \tan \chi \in [0,1),
\end{equation}
such that the metric of the \emph{round} $S^5$ becomes
\begin{equation}
\dd \Omega_5^2 = \frac{\dd x^2}{1+x^2} + \frac{x^2}{1+x^2} \, \sigma_1^2 + \frac{1}{1+x^2} \, \sigma_2^2 + \bigg( \frac{1-x^2}{1+x^2} \bigg)^2 \dd \alpha^2 + \bigg( \sigma_3 + \frac{2x}{1+x^2} \dd \alpha \bigg)^2.
\end{equation}

Finally, in these coordinates we note that the boundary condition for $G_3$ (given in \eqref{eq:G3PS}) corresponds to the boundary condition for the potential 
\begin{equation}
\label{eq:asymptoticA2}
\begin{split}
r^{-4} S_2 = \frac{e^{i\alpha}}{r (1+x^2)^{3/2} } &\Big[ - 4 x\, \mathrm{d} \alpha \wedge \left( \sigma_2 - \,i x\, \sigma_1 \right) + i\, \mathrm{d} x \wedge \left( \sigma_2 + i x\, \sigma_1 \right) \\
& \qquad + \left( (3+x^2) \sigma_2 - i \,x\, (1+3x^2) \sigma_1 \right) \wedge \sigma_3   \Big].
\end{split}
\end{equation}

%%%%% SUB-SECTION %%%%%%%%%%%%%%%%%%%%%%%
\subsection{Numerical scheme}
\noindent\indent We wish to search for black hole solutions of the coupled system of nonlinear partial differential equations (\ref{eqs:IIB}). While this can be accomplished by \emph{\`a priori} choosing a particular gauge, for instance the conformal gauge, here we will use the so called DeTurck method. This method was first introduced in \cite{Headrick:2009pv}, and recently reviewed in \cite{Dias:2015nua}. The idea is to modify the Einstein equation to the so-called Einstein-DeTurck equation:
\begin{equation}
R_{ab}-\nabla_{(a}\xi_{b)}=T^{(1)}_{ab}+T^{(3)}_{ab}+T^{(5)}_{ab}\,,
\label{eq:turck}
\end{equation}
where $\xi^a=g^{bc}\left[\Gamma_{bc}^a(g)-\Gamma_{bc}^a(g^{(0)})\right]$, $\Gamma^{a}_{bc}(\mathfrak{g})$ are the components of the Christoffel connection associated to a metric $\mathfrak{g}$, and $g^{(0)}$ is a reference metric which must have the same conformal structure as $g$, but is otherwise arbitrary. It is clear that solutions with $\xi=0$ will be solutions of the Einstein equation (\ref{eq:einstein}), but the converse might not be true. Solutions with $\xi\neq0$ are known as DeTurck solitons and play an important role in several areas of mathematics.

We want to find solutions for which $\xi=0$. In some circumstances, most notably for pure gravity under certain symmetry assumptions, it has been shown that DeTurck solitons cannot exist \cite{Figueras:2011va,Figueras:2016nmo}. Here, since we lack such theorems, we take a different route. Instead we simply solve  (\ref{eq:turck}) with appropriate boundary conditions (discussed below), and then check \emph{\`a posteriori} that $\xi$ tends to zero in the continuum limit at the expected rate dictated by the precise numerical method we decide to implement. 

The main advantage of (\ref{eq:turck}) over (\ref{eq:einstein}) is the fact that all components appear to be dynamical. That is to say, the principal symbol of (\ref{eq:turck}) is proportional to $(\mathcal{P}(g))_{ab}\equiv g^{cd}\partial_c \partial_d g_{ab}$. Since we are searching for static solutions, $\mathcal{P}=g^{cd}\partial_c \partial_d=g^{ij}\partial_i \partial_j$, where $i$ and $j$ run over spatial directions, with $g^{ij}$ a positive definite symmetric tensor. We thus conclude that $\mathcal{P}$ is an elliptic operator. Given that the problem we are solving is elliptic, we know that any two solutions cannot be arbitrarily close to each other, except at points of measure zero in the moduli space (for instance zero-modes connecting solutions in phase space), and thus a DeTurck soliton cannot be arbitrarily close to a solution of the Einstein equation. We should thus be able to differentiate DeTurck solitons from solutions of the Einstein equation by monitoring $\xi$.

Since our mass deformation will preserve the boundary metric, we will take $g^{(0)}$ to be the metric of a five-dimensional planar AdS-Schwarzschild black hole with horizon topology $\mathbb{R}^3\times S^5$. This corresponds to choosing $g^{(0)}$ to be the metric described below in \eqref{IIB:metric} with the relevant $q_i$'s given {\it everywhere} in the bulk by the constants 1 or 0 as described in \eqref{IIB:BCsUV}. We also note that we want $A_2$ to approach the Polchinski-Strassler form (\ref{eq:asymptoticA2}) asymptotically. As we shall see, our bulk metric will depend non-trivially on the $(x,y)$ coordinates, which in turn means that $A_2$ has non-zero components along these coordinates. When this happens, we also need to introduce a DeTurck-type term for the gauge potential $A_2$ in order to fix the gauge freedom $A_2\to A_2+\mathrm{d}\tilde{\chi}$, where $\tilde{\chi}$ is a regular $1-$form. There are several ways to do this (see for instance \cite{Dias:2015nua}). One possibility is to introduce a gauge-fixing term directly in (\ref{eq:g3}),
\begin{equation}
\mathrm{d}\star G_3+\star \mathrm{d}\chi-i Q\wedge \star G_3-P\wedge \star \bar{G}_3+4i G_3\wedge F_5=0,
\label{eq:a2fix}
\end{equation}
where
\begin{equation}
\chi = f(\chi_0-B\,\bar{\chi}_0)\,,\quad \chi_0 = -\star\mathrm{d}\star \Delta A_2+i_{\xi}\Delta A_2\,,\quad \Delta A_2 = A_2-A_2^{(0)}
\end{equation}
and $A_2^{(0)}$ is a reference gauge potential for $A_2$ which is required to have the same conformal structure as $A_2$, but is otherwise arbitrary. In our construction, in the sequence of the discussion that lead to \eqref{eq:G3PS} and \eqref{eq:asymptoticA2}, we will take $A_2^{(0)}$ to be {\it everywhere} in the bulk given by  $A_2$ (to be given  below in \eqref{IIB:A2f}) with $q_{i} = m$ for $i=11,\cdots,16$ and  $q_{17}=q_{18}=-\frac{2}{3}\frac{m^2}{y_+^2}(7-6\,m)x$; these boundary conditions are summarized in equation \eqref{IIB:BCsUV}. 

Both procedures detailed above produce natural gauges. For the Einstein equation, the DeTurck procedure lands on the DeTurck gauge (otherwise called Generalized Harmonic gauge) defined by $\xi=0$, and for $A_2$ we have $\chi=0$, which is a nonlinear generalization of the Lorentz gauge. However, these gauges are not imposed \emph{ab initio}, but are a bi-product of solving the Einstein-DeTurck and Maxwell-DeTurck equations. We should thus write the most general field content compatible with our symmetries.

For completeness we present our complete \emph{ansatz}:
\begin{subequations}
	\label{eqs:ansatz}
	\begin{multline}\label{IIB:metric}
	\mathrm{d}s^2 = \frac{L^2}{1-y^2}\left[-q_1\,y_+^2\,(2-y^2)y^2\mathrm{d}t^2+\frac{q_2\mathrm{d}y^2}{(2-y^2)(1-y^2)}+q_5\,y_+^2\,(\mathrm{d}w_1^2+\mathrm{d}w_2^2+\mathrm{d}w_3^2)\right]+\\
	L^2\left[\frac{q_4(\mathrm{d}x+q_3\mathrm{d}y)^2}{(1+x^2)^2}+\frac{x^2\,q_6}{1+x^2}\sigma_1^2+\frac{q_7}{1+x^2} \sigma_2^2+q_8\left(\frac{1-x^2}{1+x^2}\right)^2\mathrm{d}\alpha^2+q_9\left(\sigma_3+\frac{2\,x\,q_{10}}{1+x^2}\mathrm{d}\alpha\right)^2\right]\,,
	\end{multline}
	\begin{multline}\label{IIB:A2f}
	A_2 = L^2\,\frac{e^{i\alpha}\sqrt{1-y^2}}{y_+\,(1+x^2)^{3/2}}\Bigg[-4\,x\,q_{11}\mathrm{d}\alpha\wedge \sigma_2+4\,i\,x^2\,q_{12}\,\mathrm{d}\alpha\wedge \sigma_1+i\,q_{13}\,\mathrm{d}x\wedge\mathrm{d}\sigma_2-x\,q_{14}\,\mathrm{d}x\wedge\mathrm{d}\sigma_1\\-q_{15}(3+x^2)\,\sigma_3\wedge\sigma_2+i\,x\,(1+3x^2)\,q_{16}\sigma_3\wedge\sigma_1+q_{17}\,i\,\mathrm{d}y\wedge\sigma_2+q_{18}\,x\,\mathrm{d}y\wedge\sigma_1\Bigg]\,,
	\end{multline}
	\begin{equation}
	C_4=\frac{L^4}{4}\left[\frac{y_+^4\,y^2 (2-y^2)q_{20}}{(1-y^2)^2}\mathrm{d}t\wedge\mathrm{d}w_1\wedge\mathrm{d}w_2\wedge\mathrm{d}w_3-\frac{2\,x^2\,q_{21}}{(1+x^2)^2}\mathrm{d}\alpha\wedge\sigma_1\wedge\sigma_2\wedge\sigma_3\right]\,,
	\end{equation}
	and
	\begin{equation}
	B=e^{2\,i\,\alpha}q_{19}\,.
	\end{equation}
\end{subequations}
There are a total of 21 functions $q_i$ of two variables $(x,y)$ to be solved numerically. Note that $A_2$ and $B$ depend on $\alpha$, or in other words, transform under the $U(1)$ isometry with charges 1 and 2, respectively. However, despite the presence of this non $U(1)$-invariant matter, the energy-momentum tensor and thus the metric and $4-$form potential $C_4$ remain invariant. A similar situation is responsible for the existence of the single-Killing field black holes studied in \cite{Dias:2011at,Dias:2015rxy}. As a non-trivial verification that it is consistent for the metric and 5-form to be $U(1)$-invariant, we have checked that the equations of motion generated by  our \emph{ansatz} do close and yield a set of $20$ elliptic partial differential equations of two variables\footnote{One of the functions, namely $q_{21}$, can be removed from the system by imposing self-duality of the 5-form field strength.}. 

In these coordinates, the black hole horizon is located at $y=0$, and the constant $y_+$ appearing in (\ref{eqs:ansatz}) parameterizes the black hole temperature via
\begin{equation}
L\,T=\frac{y_+}{\pi}\,.
\end{equation}

As boundary conditions, we demand
\begin{subequations}\label{IIB:BCsUV}
	\begin{align}
	&q_{i} = 1\,\quad \text{for}\quad i=1,\,2,\,4,\,5,\,6,\,7,\,8,\,9,\,10,\,20,
	\\
	&q_{3}=q_{19}=0\,,
	\\
	&q_{i} = m\,\quad \text{for}\quad i=11,\,12,\,13,\,14,\,15,\,16,
	\\
	& q_{17}=q_{18}=-\frac{2}{3}\frac{m^2}{y_+^2}(7-6\,m)x\,,
	\end{align}
\end{subequations}
at the conformal boundary, located at $y=1$. Everywhere else in the bulk, we demand regularity.

%%%%% SUB-SECTION %%%%%%%%%%%%%%%%%%%%%%%
\subsection{Energy and a Smarr relation}
\noindent\indent In this section we will work out the relationship between the asymptotics of the solution and thermodynamical quantities. This is well-understood for asymptotically locally $AdS$ solutions, such as the 5d solutions considered in Section~\ref{sec:5d}, but is a non-trivial problem in for the full 10d geometry. In particular, we will work out the energy and derive a Smarr relation. Further details that will complement the analysis given here will be given in \cite{dias:2017a}.

First we find the asymptotic expansion of our equations of motion restricted to our symmetry. We first set
\begin{equation}
q_i(x,y)=\sum_{J=0}^{+\infty} a_i^{(J)}(x)(1-y)^J\,.
\end{equation}
Note the absence of logarithms in the expansion above, in accordance to the results of \cite{Taylor:2001pp}.\footnote{This is a non-trivial statement since, for example, for solutions of ${\cal N}=0^*$ logarithms do appear in the asymptotic expansion \cite{Taylor:2001pp}.} The energy is related to the terms up to order $J=2$. This can be easily see via dimension counting: since the stress-energy tensor has conformal dimension $4$, and asymptotically $1-y\propto z^2$ (where $z$ is the Fefferman-Graham coordinate), the relevant components for the computation of the energy appear at order $(1-y)^2\propto z^4$.

The explicit expansions of $q_1(x,y)$ and $q_{20}(x,y)$ to second order in $(1-y)$ and $q_{11}(x,y)$ to first order, the only ones we will need for our calculations below, are:
\begin{subequations}
	\begin{align}
	&q_1(x,y)=1+\frac{m^2}{y_+^2}(1-y)+\left[\alpha_0(m,y_+)-\frac{4\,m^4}{y_+^4} \left(\frac{1-x^2}{1+x^2}\right)^2\right](1-y)^2+\mathcal{O}[(1-y)^4]\,,
	\\
	&q_{11}(x,y)=m+\left[\beta_1(m,y_+)+\frac{m^2 \left(7-24 m+7 x^2\right)}{3\,y_+^2 \left(1+x^2\right)}\right](1-y)+\mathcal{O}[(1-y)^2]\,,
	\\
	&q_{20}(x,y)=1-\frac{2\,m^2}{y_+^2} (1-y)+\alpha _2(m,y_+)(1-y)^2+\mathcal{O}[(1-y)^4]\,,
	\end{align}
\end{subequations}
where $\alpha_0(m,y_+)$, $\beta_1(m,y_+)$ and $\alpha_2(m,y_+)$ are constants that dependent on $m$ and $y_+$, and will be extracted from the numerics via a fit to the numerical data.

We now turn to the calculation of thermodynamic quantities. Our analysis will closely parallel that of \cite{Dias:2017opt} (which used similar methods to study the thermodynamics of black branes in M-theory)  and will be further detailed in  \cite{dias:2017a}. A key property of our solutions is that conformal symmetry remains intact near the boundary, and therefore we expect the thermodynamic quantities to transform with a definite scaling under an overall scale transformation. The mass dimension of $m$ is $1$, which is the same as that of $T$. In addition, the entropy density has mass dimension $3$ and the energy density has mass dimension $4$. This means that if we preform the rescaling $m \rightarrow \lambda \, m$, $s \rightarrow \lambda^3 \, s$, we expect the following to be true:
\begin{equation}
\rho(\lambda^{3} s,\lambda m)=\lambda^4\rho(s,m)
\label{eq:scaling}
\end{equation}
where $s$ is the entropy density and $\rho(s,m)$ the energy density as a function of $s$ and $m$. In addition, we also expect a first law of thermodynamics of the form
\begin{equation}
\mathrm{d}\rho = T\mathrm{d}s+\vartheta \,\mathrm{d}m\,,
\label{eq:firstlaw}
\end{equation}
where $\vartheta$ is the conjugate variable to the mass deformation parameter $m$. These two equations allow us to deduce a Smarr law for our black hole solutions as follows. First, we take a derivative of (\ref{eq:scaling}) with respect to $\lambda$, and set $\lambda = 1$, yielding
$$
3 s \frac{\partial \rho}{\partial s}+m \frac{\partial \rho}{\partial m}=4 \rho(s,m)\,.
$$
We can then use the first law to express the partial derivatives in terms of temperature and $\vartheta$, yielding the Smarr relation
\begin{equation}
T\,s = \frac{4}{3}\rho-\frac{1}{3}\vartheta\,m\,.
\label{eq:smarr}
\end{equation}

In a forthcoming publication \cite{dias:2017a} we will show that one can associate a variation (in moduli space) of a conserved quantity to each Killing vector field $\hat{\xi}$ as follows. First, we can construct a closed $8-$form, which we coin $\omega_{\hat{\xi}}$, given by
\begin{equation}
\omega_{\hat{\xi}}= \omega_{\hat{\xi}}^{g}+\omega_{\hat{\xi}}^{C_4}+\omega_{\hat{\xi}}^{A_2}+\omega_{\hat{\xi}}^{\bar{A}_2}+\omega_{\hat{\xi}}^{B}+\omega_{\hat{\xi}}^{\bar{B}}\,,
\end{equation}
where
\begin{subequations}
	\begin{align}
	&h_{ab}\equiv\delta g_{ab}\,,\quad a_2\equiv\delta A_2\,,\quad b\equiv \delta B\,,
	\\
	& q\equiv f\star\left(\bar{G}_3-\bar{B}G_3\right)-i F_5\wedge \bar{A}_2+2 i C_4\wedge \bar{F}_3\,,
	\\
	&\Theta^B_{\hat{\xi}} = -\frac{f^2}{\kappa_{10}^2}(\star \bar{P}) b\,,\quad \Theta^{C_4}_{\hat{\xi}} = -\frac{4}{\kappa_{10}^2}F_5\wedge c_4\,,\quad \Theta^{A_2}_{\hat{\xi}} = -\frac{1}{4\kappa_{10}^2}q\wedge a_2\,,
	\\
	& Q^{A_2}_{\hat{\xi}} = \frac{1}{4\,\kappa_{10}^2}q \wedge i_{\hat{\xi}}A_2\,,\quad Q^{C_4}_{\hat{\xi}} = \frac{4}{\kappa_{10}^2}F_5 \wedge i_{\hat{\xi}}C_4\,,
	\\
	&\omega_{\hat{\xi}}^{g} =\frac{1}{2\kappa_{10}^2}\left[-\delta \mathrm{d}\hat{\xi}-i_{\hat{\xi}}\star [(\nabla_b h^{b}_{\phantom{b}a}-\nabla_a h)e^a]\right]
	\\
	&\omega_{\hat{\xi}}^{B} = -i_{\hat{\xi}} \Theta^B_{\hat{\xi}}\,,\quad \omega_{\hat{\xi}}^{C_4} = \delta Q^{C^4}_{\hat{\xi}}-i_{\hat{\xi}} \Theta^{C_4}_{\hat{\xi}}\quad \text{and}\quad\omega_{\hat{\xi}}^{A_2} = \delta Q^{A_2}_{\hat{\xi}}-i_{\hat{\xi}} \Theta^{A_2}_{\hat{\xi}}\,.
	\end{align}
\end{subequations}
In the expressions above, $\delta$ represents a variation along the moduli space of black hole solutions. For our specific example $\delta = \partial_m+\partial_{y_+}$. One can explicitly check, using the equations of motion (\ref{eqs:IIB}) and the fact that $\hat{\xi}$ is a Killing vector, that $\omega_{\hat{\xi}}$ is a closed $8-$form, that is to say $\mathrm{d}\omega_{\hat{\xi}}=0$.

Our \emph{ansatz} (\ref{eqs:ansatz}) is written in coordinates that are well adapted to our Killing symmetries, namely time translations $\hat{\xi}=\partial_t$ and spatial translations $\hat{\xi}_i = \partial_{w_i}$ for $i\in\{1,2,3\}$. We can thus easily construct constant $t$ and $w_i$ slices which we identify as $\Sigma_{\hat{\xi}}$. Integrating $\mathrm{d}\omega_{\hat{\xi}}$ on such surfaces gives
\begin{equation}
\label{eq:smarrdiff}
0=\int_{\Sigma_{\hat{\xi}}} \mathrm{d} \omega_{\hat{\xi}} = \int_{S^{y=1}_{\hat{\xi}}} \omega_{\hat{\xi}}-\int_{S^{y=0}_{\hat{\xi}}} \omega_{\hat{\xi}}\quad\qquad \Rightarrow \quad\int_{S^{y=1}_{\hat{\xi}}} \omega_{\hat{\xi}}=\int_{S^{y=0}_{\hat{\xi}}}\omega_{\hat{\xi}}\,,
\end{equation}
where $S^y_{\Sigma_{\hat{\xi}}}$ is a constant $y$ slice of $\Sigma_{\hat{\xi}}$ and we used Stokes' theorem. In order to render the integrals in time finite, we integrate over a period of time $\Delta t$. For integrations in $w_i$, we take $w^i$ to be periodic with period $\Delta w_i$.

Using our horizon boundary conditions, we can evaluate the right hand side of (\ref{eq:smarrdiff}) for $\hat{\xi}=\partial_t$, to find
\begin{equation}
\frac{1}{\Delta w_1\Delta w_2\Delta w_3}\int_{S^{y=0}_{\hat{\xi}}}\omega_{\hat{\xi}} = T \delta s\,.
\label{eq:bigone}
\end{equation}
This means, using our first law (\ref{eq:firstlaw}), that whatever is on the left hand side of (\ref{eq:smarrdiff}) must be given by
\begin{equation}
\int_{S^{y=1}_{\hat{\xi}}}\omega_{\hat{\xi}} = \mathrm{d}\rho-\vartheta \,\mathrm{d}m\,.
\end{equation}

Finally we note that if we look at the difference
\begin{equation}
\frac{1}{\Delta w_1\Delta w_2\Delta w_3}\int_{S^{y=1}_{\partial_t}}\omega_{\hat{\xi}}-\frac{1}{\Delta t}\int_{S^{y=1}_{\partial_{w_i}}}\omega_{\hat{\xi}} =\frac{1}{\Delta w_1\Delta w_2\Delta w_3}\int_{S^{y=0}_{\partial_t}}\omega_{\hat{\xi}}-\frac{1}{\Delta t}\int_{S^{y=0}_{\partial_{w_i}}}\omega_{\hat{\xi}} = \delta (T s)\,,
\end{equation}
which implies, via the Smarr relation \eqref{eq:smarr}, that
\begin{equation}
\frac{1}{\Delta w_1\Delta w_2\Delta w_3}\int_{S^{y=1}_{\partial_t}}\omega_{\hat{\xi}}-\frac{1}{\Delta t}\int_{S^{y=1}_{\partial_{w_i}}}\omega_{\hat{\xi}} =\delta \left(\frac{4}{3}\rho-\frac{1}{3}\vartheta\,m\right)\,.
\label{eq:bigtwo}
\end{equation}

The left hand side of both (\ref{eq:bigone}) and  (\ref{eq:bigtwo}) can be readily evaluated in terms of our asymptotic quantities for  the fields $q_i(y)$. Furthermore, they provide two differential equations for two unknowns $\rho$ and $\vartheta$, which can be solved up to two constants of integration $\rho_0$ and $C_1$:
\begin{subequations}
	\begin{equation}
	\vartheta = -\frac{3 N^2}{2 \pi ^2} \left[y_+^2\,\beta _1(m,y_+)+C_1 m^3+\frac{14 m^2}{9}\right]\,,
	\end{equation}
	\begin{multline}
	\rho = -\frac{N^2}{48 \pi ^2} \Big[6 \,y_+^4\alpha _0(m,y_+)+18 C_1 m^4+3 m^4+28 m^3+3 m^2 y_+^2+\\
	18 \beta _1(m,y_+) m\,y_+^2-18\,y_+^4+\rho_0\Big]\,.
	\label{eq:energy}
	\end{multline}
\end{subequations}

The constants $\rho_0$ and $C_1$ are fixed via supersymmetry. In a nutshell, we demand the confined phase of the theory, which is supersymmetric, to have zero energy, so that our energy measures a relative energy between our deconfined phase and the supersymmetric phase. It turns out that this is equivalent to choosing $C_1 = -4/3$ and $\rho_0=0$ in  (\ref{eq:energy}). In the expressions above, we have also used $\kappa_{10}^2 \equiv 8 \pi G_{10}= 4 \pi ^5 L^8/N^2$ with $L=1$.

The energy and entropy densities are $\rho$ and $s$. However, neither of these are invariant under conformal scalings. To fix this, we will be plotting $\widehat{\rho}=\rho/T^4$ and $\widehat{s}=s/T^3$. These scalings ensure that both $\widehat{\rho}$ and $\widehat{s}$ are a function of $m/T$ only. We can also construct a dimensionless free energy from $\widehat{\rho}$ and $\widehat{s}$ via $\widehat{f}=\widehat{\rho}-\widehat{s}$. This in turn allows us to deduce a first law for $\widehat{f}$ as follows
\begin{equation}
\left.\frac{\partial f}{\partial T}\right|_{m} = -s \quad \Rightarrow \quad \frac{\partial}{\partial T}\left[T^4 \widehat{f}\left(\frac{m}{T}\right)\right]=-T^3 \widehat{s}\left(\frac{m}{T}\right)\Rightarrow 4\widehat{f}(\widehat{m})+\widehat{s}(\widehat{m})=\widehat{m}\,\widehat{f}^\prime(\widehat{m})
\end{equation}
where we defined the conformally-invariant mass deformation $\widehat{m}\equiv m/T$, as before. All our data satisfies this form of the first law to $0.2\%$.

%%%%% SUB-SECTION %%%%%%%%%%%%%%%%%%%%%%%
\subsection{IIB description of the deconfined high-temperature phase: results \label{sec:IIBresults}}
\noindent\indent In order to solve for these $20$ functions, we discretise the partial differential equations using a pseudo-spectral collocation scheme on two orthogonal Chebyshev grids on the unit square $(x,y)\in[0,1]\times[0,1]$. We solve the resulting nonlinear algebraic equations using a standard Newton's method algorithm (see {\it e.g.} \cite{Dias:2015nua} for a detailed description of these methods).

Before proceeding, let us recall the work of Freedman and Minahan, \cite{Freedman:2000xb}, who constructed the $d=10$ black brane solution to second order in $\widehat{m}$. In particular, they made a prediction for the entropy of such solutions which, after changing to our conventions, becomes\footnote{The results in \cite{Freedman:2000xb} were left as a function of the following integral
$$
I_1\equiv \int_0^1\mathrm{d}x\frac{x^2}{(1-x^2)^2}\left[P_{-1/2}(x)-x P_{1/2}(x)\right]^2\,,
$$
where $P_{n}(x)$ is a Legendre function. We managed to compute this integral analytically. For that we first use the relation $P_{-1/2}(x)-x P_{1/2}(x)=\frac{3}{2} (1-x) \, _2F_1\left(-\frac{1}{2},\frac{3}{2};2;\frac{1-x}{2}\right)$, where $_2F_1$ is an hypergeometric function, and then we use $_2 F_1(a,b,c,z)=\frac{\Gamma (c)}{\Gamma (b) \Gamma (c-b)}\int_0^1 \mathrm{d}t \,t^{b - 1} (1 - t)^{c - b - 1} (1 - t z)^{-a}$. Once the dust settles, we find $I_1=3/2-4/\pi$.}
\begin{equation}
\widehat{s}=\frac{\pi ^2}{2}-\frac{3}{\pi ^2} \Gamma \left(\frac{3}{4}\right)^4\widehat{m}^2+\mathcal{O}(\widehat{m}^4)\,.
\label{eq:predi}
\end{equation}

\begin{figure}[ht]
	\centering
	\includegraphics[height=0.65\linewidth]{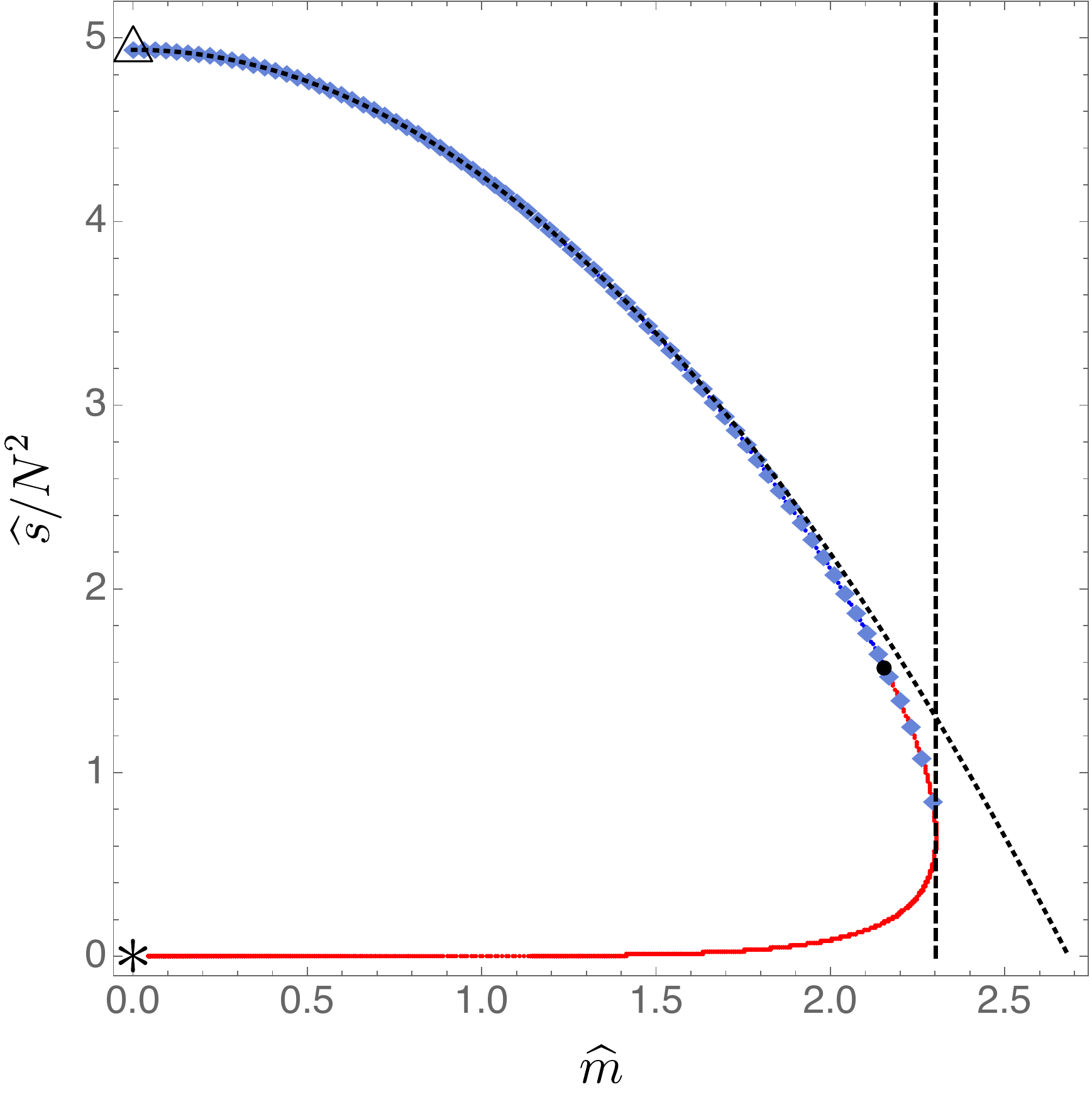}
	\caption{The conformally-invariant  entropy density $\widehat{s}$ as a function of the conformally-invariant mass deformation $\widehat{m}$: comparing the $d=5$ and $d=10$ supergravity results. The (red continuous line) $d=5$ results were already shown in Fig.~\ref{fig:thermo1}, but here we also add the IIB results (blue diamond curve which is on the top of 5d results), and we also plot the second-order (in $\widehat{m}$) perturbative IIB result (\ref{eq:predi}) depicted here as a dashed black line. We use the same color and symbol codes as in the captions of previous Figs.~\ref{fig:thermo1}--\ref{fig:thermo5b}. Namely, the vertical dashed line intersects the turning point at $\widehat{m}=\widehat{m}_{t}\sim 2.30$; see \eqref{turningPoint}. The triangle $\triangle$ (with $\widehat{m}=0$, $\widehat{\rho}\neq 0$ and $\widehat{s}\neq 0$) corresponds to the planar AdS black hole. The asterisk $\ast$ (with $\widehat{m}=0$, $\widehat{\rho}=0$ and $\widehat{s}=0$) describes a singular solution whose Weyl tensor blows up (see Fig.~\ref{fig:thermo2}); we give strong evidence that it describes the GPPZ solution (\ref{eq:super1}) or, equivalently, its IIB uplift given in \cite{Petrini:2018pjk,Bobev:2018eer}. The black disk $\bullet$ pinpoints the critical mass deformation $\widehat{m}=\widehat{m}_c\sim 2.15$, see \eqref{phaseTransition}, at which a (most likely first-order) phase transition occurs (see Fig.~\ref{fig:canonical}). The branch of solutions from $\bullet$ into $\triangle$ dominates the canonical ensemble  while the branch from $\bullet$ into $\ast$ has positive free energy density: see Fig.~\ref{fig:canonical}.}
	\label{fig:comparing}
\end{figure}

We are now in position to compare our results obtained in $d=5$ with those in $d=10$. In Fig.~\ref{fig:comparing} we show the dimensionless entropy density as a function of the mass deformation obtained from solving directly the type IIB equations of motion (represented by the blue diamonds) as well as the dimensionless entropy density obtained directly from $d=5$ (represented by the blue and red disks already displayed in Fig.~\ref{fig:thermo1}). The agreement is excellent for the parameter range where the two solutions overlap (more precisely, the parameter range where we have managed to generate the $d=10$ solution). We have also included the perturbative Freedman-Minahan result (\ref{eq:predi}) as a black dotted line. We see that, for $\widehat{m}\lesssim1$, the perturbative result (\ref{eq:predi}) gives an excellent approximation to the exact numerical result.

\begin{figure}[ht]
	\centering
	\includegraphics[height=0.45\linewidth]{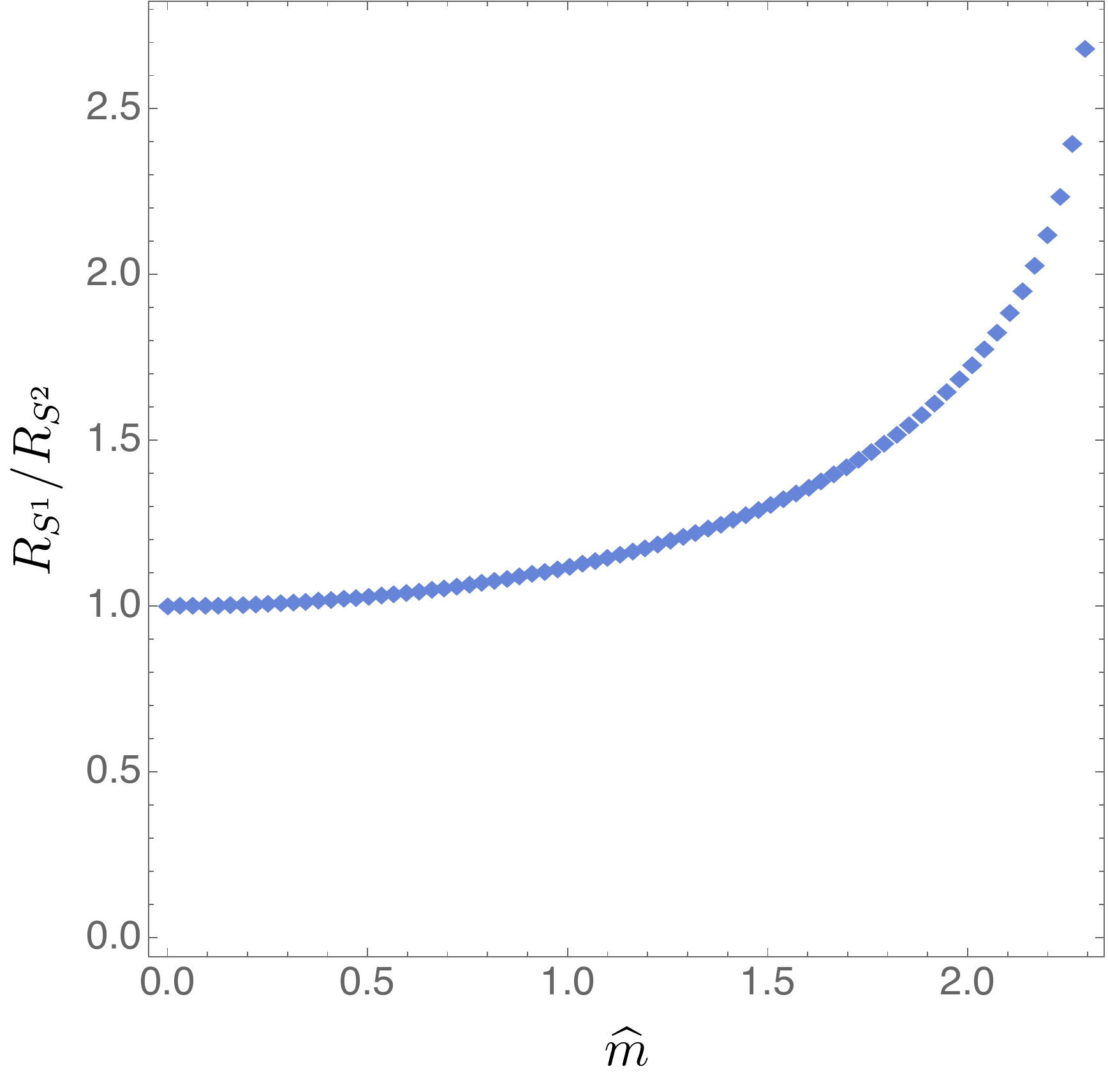}
	\caption{The ratio of the maximum of the $S^1$ (located at $x=1$ in our coordinates) and the maximum radius of the $S^2$ (located at $x=0$ in our coordinates) at the horizon as a function of $\widehat{m}$.  This seems to suggest that the solution might develop a Gregory-Laflamme instability as $\widehat{m}$ grows.}
	\label{fig:ratio}
\end{figure}

Having obtained  the ten-dimensional description of the high-temperature confined vacuum solution, we can investigate what is the distortion of the $S^5$ as $\widehat{m}$ increases. While the solution does not become singular for any value of $\widehat{m}$ over which we have managed to construct the 10d solution, the $S^5$ gets very distorted. This is best illustrated by plotting the the ratio of the maximum of the $S^1$ (located at $x=1$ in our coordinates) and the maximum radius of the $S^2$ (located at $x=0$ in our coordinates) at the horizon. This is given by the ratio $R_{S^1}/R_{S^2}=\sqrt{q_8(0,0)/q_6(1,0)}$, which we plot in Fig.~\ref{fig:ratio}. The observed hierarchy of scales suggests the existence of a kind of Gregory-Laflamme instability \cite{Gregory:1993vy} along the $S^5$, leading to the existence of a new family of black holes more distorted than the ones presented here, which would branch from the onset of the aforementioned instability. This is akin of what happens for non-uniform strings \cite{Wiseman:2002zc,Kalisch:2015via,Kalisch:2016fkm} and the lumpy holes of \cite{Dias:2014cia,Emparan:2014pra,Dias:2015pda}. Following one such solution\footnote{Note that more than one solution can bifurcate at a given onset, see for instance \cite{Dias:2014cia,Emparan:2014pra}.} will likely lead to a transition between a black hole with spatial horizon topology $\mathbb{R}^3\times S^5$ and a black hole with spatial horizon topology $\mathbb{R}^3\times S^2\times S^3$ $-$ a black ringoid.\footnote{This situation is very similar to the one studied in Ref.~\cite{Dias:2017opt}, where a smeared black brane family of solutions appeared to terminate in a singularity that suggested a new ringoid phase. One key difference is that in Ref.~\cite{Dias:2017opt} the solution became singular as the difference in free energy between the confined and deconfined phases vanished, whereas here the phase transition occurs well before the singularity.} Note that if such a black ringoid phase turns out to exist for $\widehat{m}<\widehat{m}_c$ (and $\widehat{f}<0$) then the phase transition we found at $\widehat{m}=\widehat{m}_c$ (see Fig.~\ref{fig:canonical}) will not be the confinement/deconfinement phase transition, but rather a second-order phase transition between two deconfined phases. Such analysis is outside the scope of this manuscript, but certainly merits further investigation.

%%%%%%%%%%%%%%
\section{Conclusion}
%%%%%%%%%%%%%%

Understanding and describing the confinement physics that occurs in QCD at low energies is one of the major open questions in theoretical physics. Lattice QCD is providing a path in this direction but at a slow pace and with its own limitations. Holography on the other hand is tailored to deal with the associated strongly-coupled physics but, typically, it is under good computational control only for theories that are not sufficiently similar to QCD. 

In this context, the mass-deformed ${\cal N}=4$ Super-Yang-Mills theory (also known as ${\cal N}=1^*$) definitely deserves the highest attention. For it is probably the closest theory to QCD that can be  studied simultaneously using holographic and Lattice QCD complementary methods in  user-friendly computational environments. In particular, this theory has confined and deconfined phases and associated phase transitions. That is, it has supersymmetric (confined, Higgs and oblique) vacua but also high-temperature deconfined vacua. Unfortunately, the supersymmetric vacua of the theory have not yet been constructed in the dual supergravity theory. The only supersymmetric solution that has been found  $-$  the GPPZ solution \cite{Girardello:1999bd} $-$ is singular both from the five-dimensional perspective \cite{Girardello:1999bd} and in its type IIB uplifted form \cite{Petrini:2018pjk,Bobev:2018eer}. On the other hand, the supersymmetric vacua corresponding to D3 branes polarized into D5, NS5 and (p,q)-5 branes that were argued to exist in the mean-field analysis of \cite{Polchinski:2000uf} have not yet been constructed. It seems by now clear that the supersymmetric vacua of interest are dual to cohomogeneity-three solutions, which has been the main obstacle hampering their discovery. 

However, at least the simplest high-temperature deconfined phase of this theory corresponds to a supergravity solution that has lower cohomegeneity. Therefore, in this paper we have managed to construct this deconfined phase at full nonlinear level.  We did this both in a five-dimensional supergravity truncation and by directly solving type IIB supergravity equations. For values of the conformally-invariant mass deformation $\widehat{m}=m/(\sqrt{3}T)$ smaller than $\widehat{m}_t\sim 2.30$ we have found two branches of thermal solutions with vanishing gaugino condensate. One of them agrees with the Freedman-Minahan second-order-perturbative solution \cite{Freedman:2000xb} in the small $\widehat{m}$ regime. The other one approaches the GPPZ supersymmetric solution  \cite{Girardello:1999bd,Petrini:2018pjk,Bobev:2018eer} as $\widehat{m}\to 0$.    These conclusions are best summarized in  Fig.~\ref{fig:comparing}.

We studied in detail the thermodynamic/physical properties of this deconfined solution. In particular, we found the critical temperature above which this thermal solution dominates the canonical ensemble phase diagram over the supersymmetric vacua: see Fig.~\ref{fig:canonical}. This critical temperature can or should now be reproduced by a Lattice QCD-like simulation in what would be the highest non-trivial four-dimensional precision holography test to date.      
      
Our work opens the door for interesting future research. Strictly speaking we cannot yet state whether the phase transition we find is a first-order phase transition between our  high-temperature vacuum and the confined, screening or oblique vacua of \cite{Polchinski:2000uf}, or whether there exist intermediate phases that dominate the ensemble for certain ranges of temperature. Indeed, we searched for solutions with $SO(3)$ symmetry that only have non-vanishing mass deformation and/or gaugino condensate, but there might exist other high-temperature solutions (with or without $SO(3)$ invariance) that could eventually dominate the canonical ensemble at least for some range of temperature. Some of these solutions might not even be captured by the five-dimensional description of the system. As an example, in the ten-dimensional description we have found evidence for the existence of `ringoid' thermal phases with spatial horizon topology $\mathbb{R}^3\times S^2\times S^3$. There is certainly room for such solutions to exist, and now that we have constructed the first vacuum of $\mathcal{N} = 1^*$ theory we can search for other possible thermal phases.  Ultimately, we can and intend to also search (numerically) for the cohomegeneity-3 solutions that describe the supersymmetric vacua of the theory.    

%%%%%%%%%%%%%%%%%%%%%%%%%%%%%%
\section*{Acknowledgements} 

We would like to thank Nikolay Bobev and Fri{\dh}rik Gautason for useful discussions.  Some computations were performed on the COSMOS Shared Memory system at DAMTP, University of Cambridge operated on behalf of the STFC DiRAC HPC Facility and funded by BIS National E-infrastructure capital grant ST/J005673/1 and STFC grants ST/H008586/1, ST/K00333X/1.  O.J.C.D. is supported by the STFC Ernest Rutherford grants ST/K005391/1 and ST/M004147/1 (G.S.H. also acknowledges support from the latter), and by the STFC `Particle Physics Grants Panel (PPGP) 2016' grant ST/P000711/1. I.B. was supported by the ANR grant Black-dS-String ANR-16-CE31-0004 and by the John Templeton Foundation Grant 48222. J.E.S. was supported in part by STFC grants PHY-1504541 and ST/P000681/1.  B.E.N. has been supported during the course of this work variously by ERC grant ERC-2011-StG 279363-HiDGR, and by ERC grant ERC-2013-CoG 616732-HoloQosmos, and by the FWO and European Union's Horizon 2020 research and innovation program under the Marie Sk{\l}odowska-Curie grant agreement No. 665501.

%%%%%%%%%%%%%%%%%%%%%%%%%%%%%%%%%%%
% BIBLIOGRAPHY
%%%%%%%%%%%%%%%%%%%%%%%%%%%%%%%%%%%
\bibliographystyle{utphys}
\bibliography{masterrefs}

\end{document}